\documentclass[twocolumn]{aastex63}

\usepackage{hyperref}
\usepackage{amsthm, bm}
\usepackage{tabularx}
\usepackage{array}
\usepackage{graphicx}	
\usepackage{amssymb}	
\usepackage{float}

\newcommand{\Msun}{M$_{\odot}$}
\newcommand{\latte}{$Latte$}

\begin{document}

\correspondingauthor{Nicol\'as Garavito-Camargo}
\email{ngaravito@flatironinstitute.org}

\author[0000-0001-7107-1744]{Nicol\'as Garavito-Camargo}
\affiliation{Center for Computational Astrophysics, Flatiron Institute, 162 5th Ave, New York, NY 10010, USA}

\author[0000-0003-0872-7098]{Adrian M. Price-Whelan}
\affiliation{Center for Computational Astrophysics, Flatiron Institute, 162 5th Ave, New York, NY 10010, USA}

\author[0000-0003-0872-7098]{Jenna Samuel}
\altaffiliation{NSF Astronomy and Astrophysics Postdoctoral Fellow}
\affiliation{Department of Astronomy, The University of Texas at Austin, 2515 Speedway, Stop C1400, Austin, TX 78712, USA}

\author[0000-0002-6993-0826]{Emily C. Cunningham}
\altaffiliation{NASA Hubble Fellow}
\affiliation{Department of Astronomy, Columbia University, 550 West 120th Street, New York, NY, 10027, USA}
\affiliation{Center for Computational Astrophysics, Flatiron Institute, 162 5th Ave, New York, NY 10010, USA}

\author[0000-0003-0872-7098]{Ekta Patel}
\affiliation{Department of Astronomy, University of California, Berkeley, 501 Campbell Hall, Berkeley, CA, 94720, USA} 

\author[0000-0003-0603-8942]{Andrew Wetzel}
\affiliation{Department of Physics \& Astronomy, University of California, Davis, CA 95616, USA}

\author[0000-0003-0872-7098]{Kathryn V. Johnston}
\affiliation{Department of Astronomy, Columbia University, 550 West 120th Street, New York, NY, 10027, USA}

\author[0000-0002-8354-7356]{Arpit Arora}
\affiliation{Department of Physics \& Astronomy, University of Pennsylvania, 209 S 33rd St, Philadelphia, PA 19104, USA}

\author[0000-0003-3939-3297]{Robyn E. Sanderson}
\affiliation{Department of Physics \& Astronomy, University of Pennsylvania, 209 S 33rd St, Philadelphia, PA 19104, USA}

\author[0000-0003-0872-7098]{Lehman Garrison}
\affiliation{Center for Computational Astrophysics, Flatiron Institute, 162 5th Ave, New York, NY 10010, USA}

\author[0000-0003-0872-7098]{Danny Horta}
\affiliation{Center for Computational Astrophysics, Flatiron Institute, 162 5th Ave, New York, NY 10010, USA}



\title{On the co--rotation of Milky Way satellites: \\
LMC--mass satellites induce \textit{apparent} motions in outer halo tracers}





\begin{abstract}
Understanding the physical mechanism behind the formation of a co-rotating thin plane of
satellite galaxies, like the one observed around the Milky Way (MW), has been challenging. 
The perturbations induced by a massive satellite galaxy, like the Large Magellanic Cloud (LMC) 
provide valuable insight into this problem. The LMC induces an apparent co--rotating motion in
the outer halo by displacing the inner regions of the halo with respect to the outer halo. 
Using the \latte{} suite of FIRE-2 cosmological simulations of MW-mass galaxies, we confirm 
that the \textit{apparent} motion of the outer halo induced by the infall of a massive 
satellite changes the observed distribution of orbital poles of outer-halo tracers, 
including  satellites. We quantify the changes in the distribution 
of orbital poles using the two-point angular correlation function and find that all 
satellites induce changes. However, the most massive satellites with pericentric passages 
between $\approx30-100$kpc induce the largest changes. The best LMC-like satellite analog shows the largest change in orbital pole
distribution. The dispersion of orbital poles decreases by
20$^{\circ}$ during the first two pericentric passages. Even when excluding the satellites
brought in with the LMC-like satellite, there is clustering 
of orbital poles. These results suggest that in the MW, the recent pericentric passage of 
the LMC should have changed the observed distribution of orbital poles of all other 
satellites. Therefore, studies of kinematically-coherent planes of satellites that seek 
to place the MW in a cosmological context should account for the existence of a massive 
satellite like the LMC.
\end{abstract}

\section{Introduction}\label{sec:intro}

The existence of a thin ($\approx$20 kpc) co-rotating plane of satellite galaxies around 
the Milky Way (MW) remains a potential conundrum in our understanding of how galaxies assemble over 
time. The satellites associated with this structure orbit in a planar configuration perpendicular to the plane of the disk of the MW \citep{Li21}, forming the Vast Polar Structure (hereafter VPOS) \citep{Pawlowski12}. Moreover 
in external galaxies of the local volume, distance measurements along with line-of-sight 
velocities have revealed putative planar configurations with kinematic coherence in Andromeda 
(M31) \citep{Ibata13, Sohn20} and Centaurus A (Cen A) \citep{Muller18}. Although, with the new 
measurement of M31's distance, the satellite distribution is also lop-sided \citep{Savino22}. For a comprehensive review of the history of the observations 
and proposed solutions for satellite plane formation, we refer interested readers to 
\cite{Pawlowski18, Pawlowski21, Pawlowski21b}. 

In cosmological simulations, it is rare (in an absolute sense) to find such a thin disk of
satellites co-rotating like the VPOS. For example, \cite{Pawlowski20} 
found that less than 0.1\% of the MW analogs in the Illustris simulation 
have co-rotating planes similar to the VPOS at redshift zero
(though statements about the relative occurrence rate of planes will, in general, depend on the exact definition of and identification of ``MW analog"). 
Such discrepancy is often posed as evidence of a small-scale problem for the 
$\Lambda$CDM model, ``the planes of satellites tension" \citep[e.g.,][]{Sales22}.  
However, there have been recent results from cosmological simulations that 
report that planes are rare but still consistent with $\Lambda$CDM. 
\cite{Sawala22} and \cite{Pham22} showed that in cosmological simulations that 
account for artificial tidal disruption \citep{vandenbosch18a, vandenbosch18b}, the 
probability of finding co-rotating planes similar to the VPOS is $\approx 0.3-0.5\%$,
which is within the 2-3$\sigma$ of what cosmological simulations predict. Yet, questions such as: What are the physical mechanisms that form planes? Do all spatially thin planes co-rotate? Does the same mechanism that forms the spatial planar structure induce co-rotation? What is the lifetime of these planes? 

Several possible solutions have been proposed to explain both the
planar configuration and co-rotation, including accretion through filaments 
\citep{Libeskind11, Lovell11}, satellite group infall \citep{Li08, Donghia08, Vasiliev23LMC2, Taibi23}, tidal dwarf galaxies \citep{Hammer13, Wang20, Banik22}, and mergers \citep{Smith16}. However, results from cosmological simulations that include all of these mechanisms remain inconclusive about the origin of satellite planes, for example, \citep{Kanehisa23} found that mergers through the evolution of a galaxy have a negligible effect in the present-day distribution of satellite galaxies (see discussion in \S~\ref{sec:discussion}).

To better understand co-rotating satellite planes, it is essential to understand their instantaneous 
expected occurrence rate, as well as their expected lifetimes. With both analytical orbital integration 
of satellite galaxies \citep{Fernando17, Fernando18} and cosmological simulations \citep{Buck16}, 
co-rotating planes are found to be transient and can be easily destroyed by any perturbation from
the host DM halo. However, in the absence of major accretion events, kinematically-coherent motions 
can be sustained for long periods of time \citep{Santos-Santos23}.
  
The fact that spatial planes (not co-rotating) are more common in cosmological 
simulations than co-rotating 
planes \citep[e.g.,][]{Libeskind05, Libeskind11, Garaldi18, Samuel21} also suggests 
that the co-rotation could be caused by a different mechanism from the 
one forming the spatial plane. However, most studies of satellite planes in simulations generically select MW analogs based on halo mass, ignoring the potentially important dynamical state of the MW and its satellite population. The dynamical state of the host galaxy is likely important when studying the dynamics of satellite galaxies. The occurrence rate of co-rotating 
satellite planes in simulated galaxies with similar dynamical states compared to the MW (especially given 
recent revelations about the MW--LMC interaction \citep{Vasiliev23review}) is not known, but results from \cite{Samuel21} show that in systems with LMC-like satellites the 
dispersion of orbital poles is lower compared to systems without LMC-like satellites. 
Similarly, the occurrence of satellite planes in Local Group analogs also have been found to be rare \citep{Forero-romero18}, but not more than planes around halos in isolation \citep{Pawlowski14, Li_2022}.

The existence of a massive satellite has been proposed as a
possible explanation of the co-rotation of the MW's satellite plane \citep[][hereafter GC21]{GC21planes}. \cite{Samuel21} subsequently found that MW analogs with a massive satellite near first pericenter were nearly three times more likely to have satellites with clustered orbital poles similar to the MW satellite population. Interestingly, the 
Large Magellanic Cloud (LMC), the most massive satellite of the MW, just passed its first 
pericentric passage about the MW and hence could play an important role on the
observed kinematics of the MW satellite galaxies. Recently \cite{Vasiliev23LMC2} showed that if the LMC had a previous pericentric passage at $\geq100$~kpc, it could have brought a substantial fraction of satellite galaxies, which could explain the nature of the planes of satellites within the group infall scenario.

We structured this paper as follows: In \S~\ref{sec:sims} we describe the
simulations and in \S~\ref{sec:methods_ops} the methods that we use in this work. In \S~\ref{sec:results} we present our main results. 
In \S~\ref{sec:kinematic} we discuss what are the main kinematic perturbations induced by a satellite in the host. In \S~\ref{sec:reflex} we present the amplitude of the center-of-mass (COM) host velocity induced by the satellite. We study the effect on the distribution of orbital poles in \S~\ref{sec:all-sky}. We then 
quantify the temporal evolution of the distribution of orbital poles in \S~\ref{sec:corrfunc}. 
We discuss our results in section \S~\ref{sec:discussion}. We connect these results with global measurements of the orbital poles, such as dispersion and mean root square $\Delta_{sph}$ in \S~\ref{sec:global}.  We conclude in \S~\ref{sec:conclusions}.

\section{A review of the impact of massive satellites in the outer halo}\label{sec:kinematic}

Massive satellites perturb the dynamical state of their host galaxies, through angular momentum transfer, 
the displacement of the host--satellite system barycenter (inducing an offset in net motion of the inner 
regions with respect to the outer regions; \citealt{Salomon23COMshifts}), and through the dynamical 
friction wakes they induce in the hosts' DM halo. All of these effects can have an impact on
the observed distribution of orbital poles of outer halo tracers.
Massive satellites on eccentric orbits also experience orbital ``radialization" \citep{Amorisco17, Vasiliev22}. All of these effects 
have an impact on the observed distribution of orbital poles of outer halo tracers from an observer 
in the inner galaxy.

In this paper we follow up on the work presented
in \cite{Samuel21} and GC21. In GC21, it was shown that a 
massive satellite like the LMC can induce {\textit{apparent}} co-rotation patterns in the 
outer halo of an idealized MW--LMC like system. In a galaxy without a massive satellite perturber, the entire galaxy is in the same reference frame. 
However, when a massive satellite approaches its pericenter, the inner galaxy\footnote{For the context of this paper, inner galaxy refers to the disk and halo within $\approx 30$~kpc} can react faster to the passage of the satellite and follow its trajectory. The 
dynamical times in the outer halo are longer and hence it lags in following the massive
satellite. This results in a relative displacement between the reference frames of the 
inner halo and the outer halo. In other words, the inner galaxy is not in the same inertial 
reference frame for the outer halo in the presence of a massive satellite. 
This was recently shown in cosmological simulations of LG analogs in \cite{Salomon23COMshifts}. From now on we will refer to the displacement of positions (of DM and stellar particles) 
as \textit{collective response} and to the displacement in velocities as \textit{reflex motion} following \cite{Petersen19-disk}.

In the MW--LMC system, both the reflex motion and collective response are part of the halo response 
where DM wakes are also induced by the satellites \citep[e.g.,][]{Ogiya16, garavito-camargo19a, Tamfal21, Trelles22, Rozier22}. 
The collective response consists of several modes that are excited due to
the satellite passage, but by far the largest amplitude is in the dipole mode. As shown in \cite{Weinberg22}, 
the dipole mode is the easiest to excite and is weakly damped, persisting for several dynamical 
times. Such common and persistent perturbation should be observed in many galaxies and would provide further
evidence of the response of DM halos. 
As shown in \cite{Cunningham20, Salomon23COMshifts}, substructure within a MW analog halo can also mimic
velocity dipoles in MW halo stars. For example, halos with recent accretion events or halos accreting several  satellites less massive than the LMC show stronger dipoles than those predicted from LMC perturbations. However, in halos with a more quiescent accretion history, the dipole induced by the LMC 
is comparable in magnitude to that produced by substructure.

Thanks to the data from the {\textit{Gaia}} mission \citep{Gaia18}, a velocity offset between the inner and 
outer galaxy has now been measured and is interpreted as a \textit{reflex motion} of the inner galaxy as a 
response to the infall of the LMC \citep{Petersen20, Erkal21}. Such motion has several consequences for dynamical studies in the Galaxy. For example, integrating
orbits of any tracer in the outer halo must account for  
both the reflex motion and collective response \citep{Patel20, VasilievTango, Lilleengen23}.  
This complicates interpretations of the measurement of the shape of the MW's DM halo (GC21), which is no 
longer axisymmetric. It also biases measurements of the mass of the MW using dynamical arguments \citep{Erkal20MWLMCmass, Magnus22, Chamberlain23}. 
All these observational and theoretical works highlight the importance of taking
into account the disequilibrium state of the MW halo. 
It is therefore natural to think that the observed VPOS would also contain information about the disequilibrium state of the MW's halo (both the reflex motion and collective response), because the VPOS is measured using the positions and velocities of 
the satellite galaxies. 

If the outer halo of the MW appears to be co-rotating with the LMC (as measured from the inner halo) all the angular momentum measurements
of the outer halo would be biased. One would need to correct for the reflex motion and the collective response 
to properly interpret the VPOS signal. \cite{Pawlowski22} presented idealized N-body simulations of the MW--LMC similar to those presented in 
\cite{garavito-camargo19a}. When comparing the
observed kinematics of the MW satellites to the DM particles in the simulation, \cite{Pawlowski22} found that the effect of the LMC on the orbital poles clustering of the satellites is
negligible. Yet, some
assumptions in the idealized simulations might complicate direct comparisons with
observations. For example, in \citep{Pawlowski22} the dynamics of the MW satellite galaxies is compared to the dynamics of  the dark matter particles of  a spherical halo which do not reproduce the observed phase-space distribution of the satellites.
In addition, physical processes such as dynamical
friction and cosmological initial phase-space conditions are not taken into
account in the idealized simulations. However, these complexities can be overcome 
with zoom-in cosmological simulations, where the dynamical state of the galaxy is
more realistic and one can separately study the satellite galaxies from the dark matter. 

The goal of this paper is to study host--satellite
interactions similar to the MW--LMC interaction in a cosmological context. We use 
the \latte\ suite of FIRE-2 cosmological simulations in addition to idealized simulations to explore 1) What is 
the effect and magnitude of a massive satellite on the orbital pole 
configuration of the host halo tracers? 2) What are the time scales of 
the observed features in the orbital poles? 3) Do any massive satellites induce
co-rotation patterns in the host? In a companion paper
(Patel et. al., in preparation), we will focus on the MW--LMC satellite dynamics accounting for the
time-dependent perturbations induced by the LMC using basis function expansions.

\section{Simulations}\label{sec:sims}

\begin{table*}
    \centering
    \textbf{MW--LMC analog}
    \begin{tabular}{c c c c c c c c}
    \hline
    \textbf{\latte{} simulations} & $\bm{m12b}$ & $ m12c$ & $ m12f$ & $\bm{m12i}$ & ${m12m}$ & $m12r$ & $m12w$ \\
    \hline
    \hline
    $\rm{M_{host}}$ at $z$=0 [$\times 10^{12}$ \Msun] &  1.43& 1.35& 1.71& 1.18& 1.58& 1.1& 1.08\\

    $\rm{M_{host}}$ at Sat. infall [$\times 10^{12}$ \Msun] &  0.9& 0.6 & 1.28, 0.79 & 0.86 & 1.1 & 0.27, 0.33, 0.29 & 0.35, 0.43, 0.37 \\

    $\rm{M_{sat}}$ at infall [$\times 10^{11}$ \Msun] &  \textbf{2.1} & 1.6 & 1.5, \textbf{0.8} & 0.28 & 0.34 & 2.0, 1.3, \textbf{0.6} & 0.8, 0.5, 0.4 \\ 
    Host-Sat mass ratios at infall  &  0.23& 0.32 & 0.14, 0.11 & 0.03 & 0.04 & 0.8, 0.43, 0.22 & 0.26, 0.14, 0.125 \\
    Pericentric distance [kpc] &  \textbf{37.9} & 18.1 & 0, \textbf{35.7} & 29.5 & 77.6 & 38.7, 30, \textbf{53.4} & 7.7, 78.2, 23.1 \\
    Time of 1st pericenter [Gyr] & 8.8 & 12.9 & 7.3, 10.8  & 8.05 & 10.4 & 11.1, 13.1, 11.9 & 8, 11.4, 6.9\\
    Satellite infall times [Gyr] &  8 & 11.5 &  6.32, 9.8 &  & 6.6 & 9.7, 10.7, 11.7&  10.1, 7.4, 6.09\\
 
    \hline 
    \end{tabular}
    \caption{Properties for the hosts and massive satellite galaxies from the \latte{} simulations studied in this paper. All reported DM halo masses are $M_{200m}$, defined as the total mass within a spherical radius with a mean density of 200$\times$ the matter density of the Universe. Data adapted from \cite{Samuel21, Wetzel23}. Given the satellite and host properties, we choose $\bm{m12b}$ and $\bm{m12i}$ as the MW-LMC analog and Low-mass satellite analog respectively. Columns with more than one value pertain to multiple massive satellites of a single host. Satellites highlighted in bold are the ones that have similar pericentric distances and masses to the LMC.}
    \label{tab:fire_sats}
\end{table*}

We use the \latte{} suite of zoomed cosmological-hydrodynamical simulations, which along the FIRE-2 simulations, are publicly available \citep{Wetzel23} at \url{http://flathub.flatironinstitute.org/fire}.
These are isolated MW-like galaxies (selected to be at 
least $5\times R_{200m}$ from the nearest MW--mass halo 
at $z=0$). \latte{} used the Feedback In Realistic 
Environments (FIRE-2) physics model\footnote{\href{https://fire.northwestern.edu/}{https://fire.northwestern.edu/}}, which includes 
state-of-the-art models
for gas physics, star formation, and stellar feedback. The gas models
used include a metallicity-dependent treatment of radiative heating and cooling across $10-10^{10}$ K \citep{hopkins2018fire}, a cosmic
ultraviolet background with early HI reionization $z_{reion} \approx 10$ \citep{Faucher-Giguere09}, and an explicit model for turbulent 
diffusion of metals via turbulence \citep{Hopkins16, Su17, Escala18}. Star formation occurs in gas that is self-gravitating, Jeans unstable, cold ($t \leq 10^4$K),
dense ($n \geq $ 1000 cm$^{-3}$), and molecular (following \cite{Krumholz11}). Star particles 
represent individual stellar populations under the assumption of a Kroupa stellar initial mass function 
\citep{Kroupa01}. Once formed, star particles evolve according to stellar population models from {\textsc{starburst99 v7.0}} \citep{Leitherer99}.
FIRE-2 includes several stellar feedback processes, including core collapse and type Ia supernovae, 
continuous stellar mass loss, photoionization, photoelectric heating, and radiation pressure. The cosmological parameters used in these simulations are summarized in Table~1 of \cite{Wetzel23}. The simulations were run with the FIRE-2 model whose detailed information can be found in \citep{hopkins2018fire}. 

The suite we use consists of seven MW-mass halos with
masses at $z=0$ of $M_{200m} = 1-2 \times 10^{12}$ \Msun. The zoom-in region, which extends $\approx1-2$~Mpc around each host galaxy, has a mass resolution of 
$m_{\star} = 7070$ \Msun ~and $m_{DM} = 3.5 \times 10^4$ \Msun, which makes it possible to 
capture the dynamic processes that take place during the evolution of these 
halos. 
Halo merger trees and (sub)halo catalogs were constructed with the \textsc{Rockstar} 6-D halo finder \citep{behroozi2012rockstar}, and \textsc{Consistent-trees} \citep{behroozi2012consistent}. 
We performed
this post-processing and the remainder of our analysis using the \textsc{Gizmo Analysis} and
\textsc{Halo Analysis} software packages \citep{Wetzel20a, Wetzel20b}.

The \latte{} simulations reproduce key observational properties of Local Group galaxies. 
For satellite dwarf galaxies, it has been shown that the observed radial distribution of satellite galaxies is reproduced by the simulations \citep{Samuel20}, and the distribution of stellar masses, rotation curves, and  the velocity dispersion of these galaxies broadly agrees with that of the Local Group \citep{Wetzel16, GK19a, Santiestevan23}. Furthermore, the simulations do not suffer from the too-big-to-fail problem \citep{GK19a}. The star formation histories of field and satellite galaxies in the simulations are consistent with the variety observed in local group \citep{GK19b}. 
The stellar-to-halo mass relation of the MW analog host galaxies in the simulations is in agreement with observations \citep{hopkins2018fire}, and the masses of the MW analog stellar halos are also in agreement with observations \citep{Sanderson18}. Additional properties of the MW analog stellar halo such as chemical abundance rations were characterized by \cite{Cunningham22}. Predictions of the observable properties of accretion events in the MW analogs were explored by \cite{Horta23b}, as well as the stellar mass distribution of stellar streams \citep{Shipp22} and their progenitors \citep{Panithanpaisal21}.

These results make the \latte{} suite ideal for studying the halos of MW-like galaxies in a cosmological context. For example, 
their DM halo shapes have been quantified in \cite{Baptista23, Nondh23}, and the impact of massive satellites on the stability of action variables is presented in \cite{Arora22}. Orbital 
properties of satellite galaxies around MW--like galaxies have been presented in \cite{Santiestevan23}. The role of environmental quenching in dwarf galaxies has been studied in \citep{Samuel22, Jahn22}.  

\subsection{Idealized MW--LMC simulations:}\label{sec:mwlmc_sims}

\begin{table}
    \centering
    \begin{tabular}{c c c c c c c c c}
    \hline
    \textbf{Idealized simulations} & \textbf{MW--LMC idealized} \\
    \hline
    \hline
    $\rm{M_{host}}$ at $z$=0 [$\times 10^{12}$ \Msun] & 1.03 \\
    $\rm{M_{host}}$ at Sat. infall [$\times 10^{12}$ \Msun] & 1.03 \\
    $\rm{M_{sat}}$ at infall [$\times 10^{11}$ \Msun] & 0.8, 1, 1.8, 2.5 \\ 
    Host-Sat mass ratios at infall  & 0.08, 0.1, 0.17, 0.24  \\
    Pericentric distance [kpc] & 45 \\
    Time of 1st peri. [Gyr] & 1.5\\
    Satellite infall times [Gyr] & 11-12 \\
    \hline 
    \end{tabular}
    \caption{Properties of the four idealized simulations studied in this paper. All reported DM halo masses are $M_{200m}$, defined as the total mass within a spherical radius with a mean density 200$\times$ the matter density of the Universe.}
    \label{tab:ideal_sats}
\end{table}

In addition to the $Latte$ suite, we use the idealized N-body simulations of the MW--LMC
interaction presented in \cite{garavito-camargo19a} and \cite{GC21bfe}. The main structural properties of these simulations are summarized in Section~3.2 and in Table~1 of \cite{GC21bfe}. In Table~\ref{tab:ideal_sats} we summarized the main properties of the satellites and MW halos. We re-run these simulations
in Gadget-4 \citep{Gadget4} using the Fast Multipole Method (FMM) gravity solver to guarantee
momentum conservation \citep{FMM}. We ran the simulations for a total of 8~Gyrs to study 
the future evolution of the LMC around the MW, starting from 2~Gyrs prior to the LMC's first pericenter at $z=0$ up to 6~Gyrs in the future.

\begin{figure}
    \centering

    \includegraphics[scale=0.6]{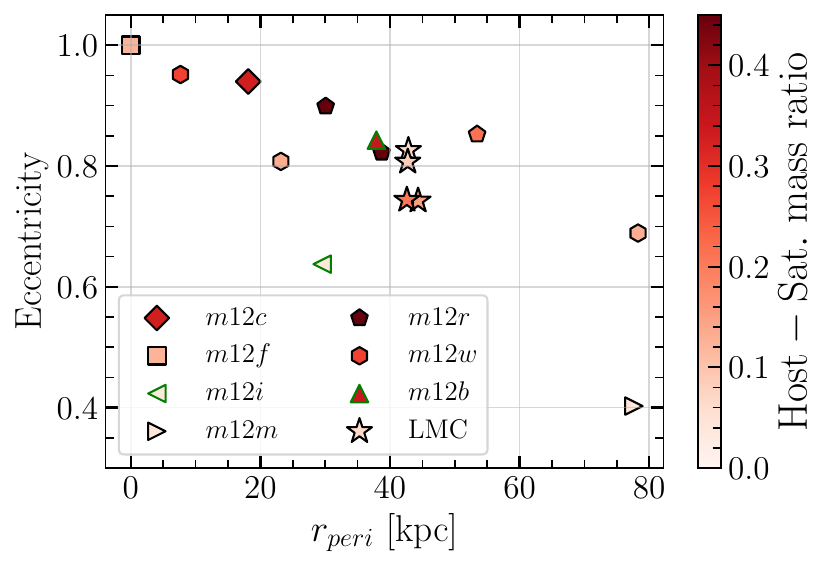}
    
    \caption{Orbital properties of the most massive satellites in the $Latte$ suite. For comparison, the MW--LMC idealized simulations are shown with star markers. The color bar shows the satellite--host mass ratio at infall. Overall the majority of the satellites in $Latte$ are on more eccentric orbits and have shorter pericentric passages. In terms of eccentricity, pericenter distance, and satellite--host mass ratios $m12b$ is the closest analog to the MW--LMC.}
    \label{fig:orbits_properties}
\end{figure}

\subsection{MW--LMC analogs in FIRE}
In this paper, we are primarily interested in understanding the distribution of orbital poles before and after the pericentric passages of massive satellites. We therefore limit our study to LMC-analog satellite systems in \latte{} by identifying the most massive accretion events 
for the 7 \latte{} halos. A summary of the main properties of these massive satellites in each halo 
can be found in Table~\ref{tab:fire_sats}. We then select satellites that have mass ratios
around 1:5 at the time of infall between $z=0-1$. 
Generally, most of the MW--LMC--like accretion events happen during this period of 
time \citep[e.g.,][]{Boylan-Kolchin11}. In addition, the host 
galaxy is undergoing steady-state star formation between $z=0-1$ 
in a thin disk as in the MW. These conditions make the \latte{} 
halos good analogs to study the impact of massive satellites like the LMC. 
Out of the 7 hosts in \latte, $m12i$, and $m12m$ do 
not accrete any massive satellites between $z=0-1$. On the other hand, $m12r$ and $m12w$ 
experience three mergers each. In the particular case of $m12r$, 
the mergers happen simultaneously and the resulting response of the DM halo is more complicated 
than in the other systems. $m12f$ experiences two accretion events, one of them being a direct head-on 
collision. $m12c$ has an LMC-like accretion event, however, the mass of the host is $1.9\times 10^{12}$M$_{\odot}$
and the satellite--host mass ratio (1:11.8) is lower than what
is expected for the MW--LMC ($\approx 1:4-5$). $m12b$ does have a 
satellite--host mass ratio of 1:4.3 and pericenter 
distance of 37.9 kpc and hence is the closest analog to the MW--LMC.
In conclusion, out of the 7 MW-like galaxies, 5 of them have a total of 10 massive satellites. Out of those,
three have similar masses and pericentric passages as the LMC highlighted in bold font in 
Table~\ref{tab:fire_sats}. Note that all the selected massive satellites overlap with the four halos 
already studied by \cite{Samuel21} with the addition of $m12r$.

In Figure~\ref{fig:orbits_properties} we show the eccentricities, pericenter distances, and 
satellite--host mass ratios at infall of all the massive satellites in \latte{}. 
We also include the four idealized MW--LMC simulations described in section~\ref{sec:mwlmc_sims}. Most of the massive satellites in \latte{} are on more eccentric orbits than the 
LMC and most of them have smaller pericenter distances. The closest MW--LMC analogs in \latte{} are the MW--LMC analog $m12b$ (triangle marker), $m12f$ (square marker), and $m12r$ (pentagon markers). 
For simplicity, we chose to only focus on two \latte{} halos $m12i$ and $m12b$. $m12i$ is representative of a halo that 
has not experienced the perturbations of an LMC-like satellite (`unperturbed halo') and $m12b$ 
will be our fiducial `MW--LMC analog' given its proximity to the LMC's eccentricity, pericenter distance, and mass ratio. The DM and stellar density of the $m12b$ system 
at pericenter is shown in Figure~\ref{fig:m12b_sketch}. A full comparison with all 7 \latte{} halos in Appendix~\ref{sec:appendix}.

\begin{figure*}
    \centering
    \text{Projection of Dark Matter (left) and stellar (right) distributions of MW--LMC analog in \latte{} ($m12b$)} \par\medskip
    \includegraphics[scale=0.6]{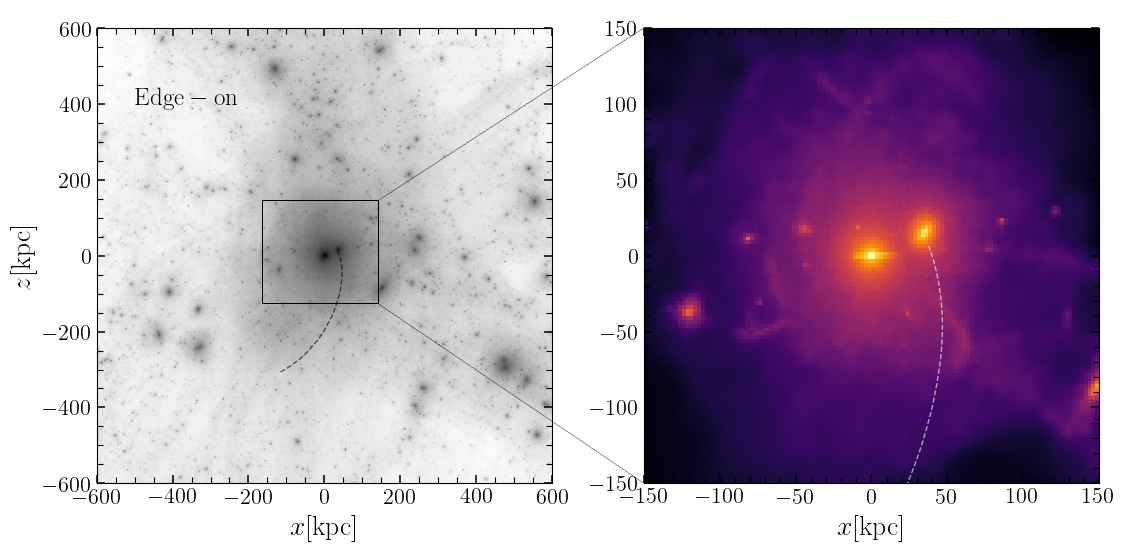}
    \caption{2D density projection of the dark matter (left panel) and star (right panel) particles of our fiducial MW--LMC analog system in the \latte\ suite, $m12b$, at the first pericenter passage (cosmic time of 8.8 Gyrs). Here we use a reference oriented such that the disk of the host galaxy is in the x--y plane through the simulation. All the coordinates (positions and velocities) are computed with respect to the center of the disk of the host galaxy. The orbital history of the LMC-like satellite is shown with dashed lines. This figure was rendered using the \textsc{pynbody} visualization module.}
    \label{fig:m12b_sketch}
\end{figure*}

\begin{figure*}[ht]
    \centering
    \includegraphics[scale=0.55]{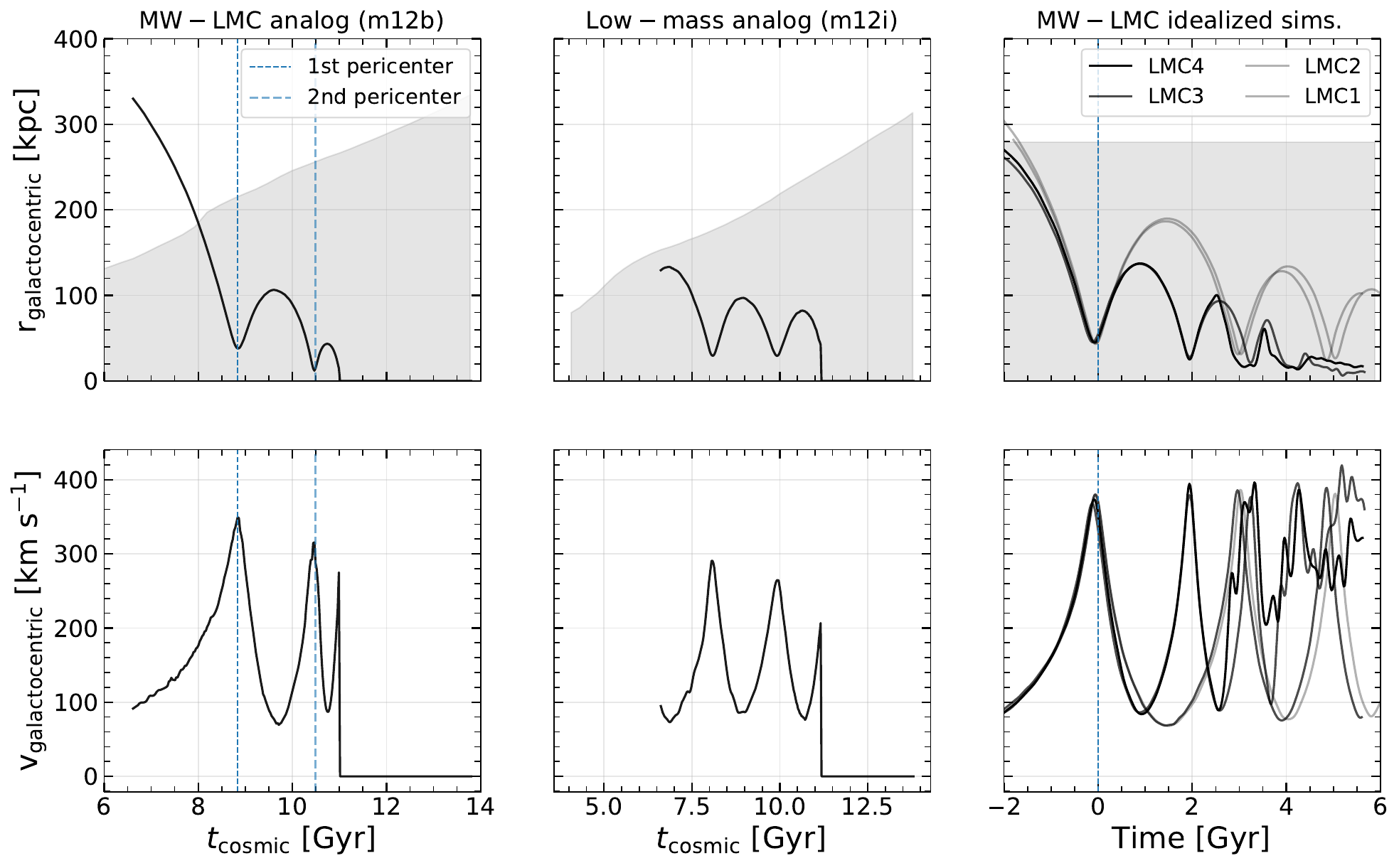}
    \caption{Each satellite's galactocentric position (top panels) and velocity (bottom panels) as a function of time. Grey-shaded regions illustrate the virial radius of the host's halo. The satellite in MW--LMC analog (left panels) is on an eccentric orbit that after two pericentric passages merges with the host.  A similar behavior is seen in the MW--LMC idealized simulations (right panels) for the most massive LMC models (LMC3 and LMC4). In the middle panel, the low-mass satellite exhibits a more circular orbit that merges with the host at the third pericentric passage.}
    \label{fig:satellite_orbits}
\end{figure*}

The orbits of the satellite galaxies in the low-mass satellite ($m12i$), the MW--LMC analog ($m12b$), and the MW--LMC idealized sims are shown 
in Figure~\ref{fig:satellite_orbits}. The velocities and positions were measured 
in the reference frame of the MW host halo as defined by the halo finder positions and velocity. 
We included the orbit of the most massive satellite in $m12i$, which is in a very circular 
orbit and produces a large stream. This satellite, however, is not massive enough to 
induce significant perturbations to the orbital poles of the host (DM, stars, satellites, or subhalos), as we will show in Section~\ref{sec:results}. The shaded regions 
illustrate the virial radius of the host galaxy as a function of time. The infall 
properties of the MW and LMC halos were taken when the satellite crossed the virial radius (see
Table~\ref{tab:fire_sats}). Pericentric passages are marked with vertical dashed blue lines. 

\section{Quantifying the angular distribution of orbital poles}\label{sec:methods_ops}

In all of our calculations, we fixed the reference frame center on the disk of the host galaxy and at a fixed longitude. 
Using the angular momentum routine implemented in \textsc{pynbody},
\footnote{\href{https://pynbody.github.io/pynbody/_modules/pynbody/analysis/angmom.html\#faceon}{https://pynbody.github.io/pynbody/\\ \_modules/pynbody/analysis/angmom.html\#faceon}}
we re-orient the halo at every snapshot so the disk of the galaxy lies on the x--y plane. 
Since the disk is not fully formed at earlier times ($z\geq 1$), there are small changes 
in the direction of the angular momentum of the disk between snapshots, which is why 
we restrict our analysis to $z\leq 1$. Since the longitude is fixed, the reference frame is not allowed to rotate along the z-axis between snapshots.

\subsection{Orbital poles metrics}\label{sec:poles_metrics}

It is common to quantify the clustering of satellite orbital poles by computing their 
dispersion ($\Delta_{\textrm{orb}}$). $\Delta_{\textrm{orb}}$ is defined as the RMS angular
 distance of the satellites' orbital angular momentum vectors 
with respect to the mean orbital angular momentum ($\hat{n}_{\textrm{orb, avg}}$) of all 
the satellites ($N_{\textrm{sat}}$):

\begin{equation}\label{eq:op_disp}
    \Delta_{orb} = \sqrt{\frac{\sum_{i=1}^{N_{sat}} [{\texttt{arccos}}({\bf{\hat{n}}_{\textrm{orb, avg}}} \cdot {\bf{\hat{n}}_{orb, i}} )]^2}{N_{\textrm{sat}}}}
\end{equation}

Similar to the dispersion of orbital poles, the spherical standard distance $\Delta_{\textrm{sph}}$ between the $k$ closest satellites or subhalos in orbital poles is often used. 
In this case, the dispersion is called $\Delta_{\textrm{sph}}$ and quantifies the clustering of $k$ orbital poles. Lower values of $\Delta_{\textrm{sph}}$ imply a lower clustering of the
k-orbital poles. We compute $\Delta_{\textrm{sph}}$ following the definition of \cite{Metz07}: 

\begin{equation}\label{eq:op_sph}
     \Delta_{sph}(k) = \sqrt{\frac{\sum_{i=1}^{k} [{\texttt{arccos}}({\bf{\hat{n}}_{orb, avg}} \cdot {\bf{\hat{n}}_{orb, i}} )]^2}{k}} 
\end{equation}

\subsection{Correlation function}\label{sec:twop_corr}

One of the main goals of this paper is to study the temporal evolution of the angular distribution of orbital poles. 
To do so, we use the two-point angular correlation function as a function of time. We use the \textit{natural estimator} \citep{Peebles99}
$\omega (\theta)$ implemented in \textsc{Corrfunc}\footnote{\href{https://corrfunc.readthedocs.io/en/master/index.html}{https:
//corrfunc.readthedocs.io/en/master/index.html}} \citep{10.1007/978-981-13-7729-7_1,corrfunc} defined as:   
\begin{equation}
    \omega(\theta) = \frac{DD}{RR} -1,
    \label{eq:corrfunc}
\end{equation}
where $RR$ is the number of any pair of a random isotropic distribution of $N$ orbital poles
in the sky. We compute $\omega(\theta)$ in angular annular bins of width 
$\Delta \cos \theta_{i} = \cos\ \theta_{i+1} - \cos\ \theta_{i}$ where $\theta_{i}$ and $\theta_{i+1}$ define the
width of the annulus. As such the number of random pairs $RR$ can be computed analytically: 
\begin{equation}
    RR = \frac{-N (N-1)}{2} \Delta \cos\ \theta.
\end{equation}

Similarly, $DD$ in Equation~\ref{eq:corrfunc} is the number of pairs in the measured distribution of orbital poles.  
Since we are interested in quantifying the evolution relative to the pre-infall distribution of orbital poles we defined $\tilde{\omega}$ to estimate the relative change in the two-point angular correlation function:
\begin{equation}
    \tilde{\omega}(\theta, t) = \frac{DD(t)}{DD(t_{\rm{infall}})} - 1 ,
\end{equation}\label{eq:relcorrfunc}
where $DD(t_{\rm{infall}})$ is the number of pairs at infall $t_{infall}\approx8$ Gyr (snapshot 300). 
Conceptually, the correlation function is the probability over random of finding a pair of orbital poles separated by an angular distance $\theta$. As such, higher positive values of $\omega$ imply higher clustering of poles, and values closer to zero imply a more isotropic distribution of poles at a given angular distance $\theta$. Clustering at small scales will be higher if $\omega$ is larger for smaller values of $\theta$. Similarly, 
higher positive values of $\tilde{\omega}$ imply enhancement of orbital poles clustering over the clustering at time $t_\mathrm{infall}$. Negative values of $\tilde{\omega}$ imply anti-correlation.

\subsection{Definitions of tracer populations}

We will characterize the distribution of orbital poles using different tracers, including DM particles, 
star particles, dark subhalos, and luminous satellite galaxies. The definition of the tracers samples are listed in Table~\ref{tab:subhalos_selection}. To identify the particles associated with the massive 
satellite we use the particle IDs provided by the halo finder at the snapshot
where the satellite was outside the virial radius of the host and where it has its peak mass. We 
remove these particles in all of the analysis presented hereafter. 

In this paper, we define the outer halo to be at distances $\geq 50$~kpc. This is a rather arbitrary choice, but it is motivated by the fact that this is approximately the distance in which the inner halo is displaced with respect to the outer halo \citep[e.g.,][]{Salomon23COMshifts}.  

\begin{table*}[ht]
    \centering
    \begin{tabular}{c c c }
         \hline
         \textbf{Definition of tracer sample} &  \textbf{Selection criteria} & Analysis used  \\
         \hline
         \hline
         Host DM & DM particles of the host excluding the  & correlation function \S~\ref{sec:corrfunc}  \\
          & particles from the massive satellite & \\
         Host stars & Star particles from the host, excluding & correlation function \S~\ref{sec:corrfunc} \\
         & the stars from the massive satellite & \\
         All subhalos &  DM subhalos within 50-300 kpc & dispersion and mean \S~\ref{sec:global} \\
         & with peak mass $\geq 10^7$\Msun & \\
         Satellites &  Satellites with within 50-300 kpc & dispersion and mean \S~\ref{sec:global} \\
         & DM halo peak mass $\geq 10^7$\Msun & \\
         Top 11 luminous satellites & 11 most massive satellites & $\Delta_{sph}(k)$ \S~\ref{sec:global} \\
         & in each halo within 300 kpc \\
         \hline
    \end{tabular}
    \caption{Selection criteria of the halo tracers in the \latte{} halos and the analysis done with each tracer.}
    \label{tab:subhalos_selection}
\end{table*}

\section{Results}\label{sec:results}

As discussed in Section~\ref{sec:kinematic}, satellite galaxies induce several perturbations in the host DM halo. 
To investigate the effect of satellites on the observed distribution of orbital poles, we start by
quantifying the reflex motion induced in the host in both idealized and cosmological zoom-in simulations 
(\S~\ref{sec:reflex}). In \S~\ref{sec:all-sky} we qualitatively show how satellites change the orbital poles distribution (of DM particles, 
satellites and subhalos of the host) at the time of the first pericentric passage. We use correlation functions in \S~\ref{sec:corrfunc} to quantify the spatial and temporal evolution of the distribution of orbital poles.

\subsection{Amplitude of the reflex motion in MW--LMC analogs}\label{sec:reflex}

As reported in \citep{GC21planes} `apparent' co-rotation 
signatures in the outer halo are produced when the inner halo 
moves with respect to the outer halo. The direction of this 
motion defines the kinematic signature. Note that the co-
rotation appears when the COM motion is not parallel to the 
velocity displacement vector (as in a linear motion). Studying 
the COM motion in cosmological halos is not trivial as the halo 
is moving in the cosmological box, and it is not obvious to 
isolate the displacement induced by a satellite. However, the 
velocities are roughly constant in the box and hence the reflex 
motion induced by the satellite is easier to characterize.

\begin{figure*}[ht]
    \centering

    \includegraphics[scale=0.55]{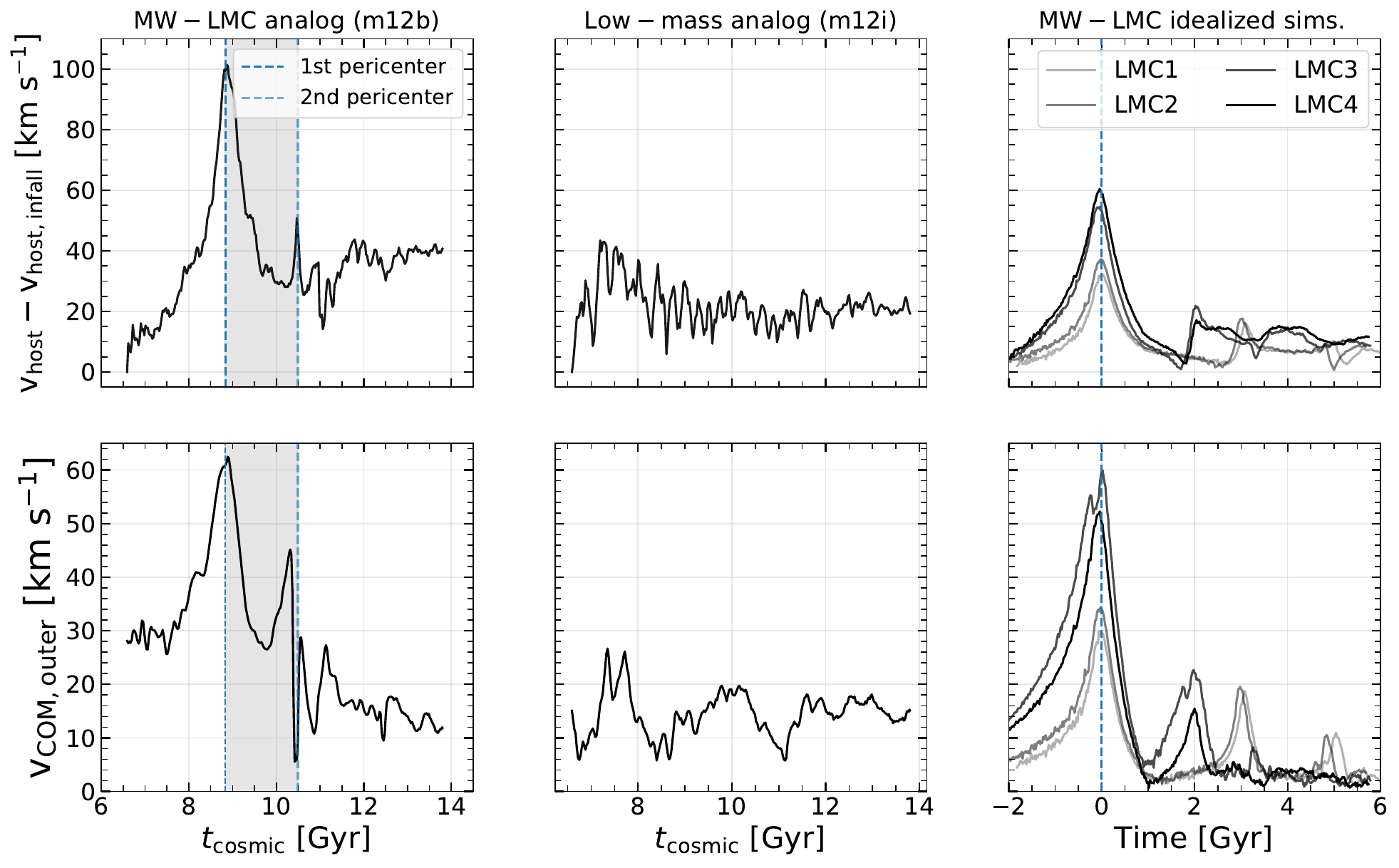}

    \caption{Magnitude of the host's velocity in the box reference frame and relative to the host's velocity at $t=6$~Gyr is plotted as a function of time (upper panels) for the MW--LMC analog 
    (left panel), the low-mass satellite analog (middle panel), and the idealized MW--LMC simulations (right panel). The host's 
    velocity increases by 80~$\rm{km\ s^{-1}}$ between infall time ($8\sim \rm{Gyr}$) and its pericenter. Finally, the velocity settles after the merger at $\approx$ -40~$\rm{km\ s^{-1}}$ relative to the velocity at $t=6$~Gyr. Such rapid motions happen in less than 2~Gyrs,
    which is comparable to the dynamical times of the MW-halo at $\approx 100$~kpc.  The lower panels show the outer halo 
    COM velocity relative to the inner halo velocity reported by the halo finder. The outer halo velocity is measured using the subhalos (black) that are beyond 
    50~kpc. In the MW--LMC analog halo ($m12b$) the outer halo already has a net motion of 30~$\rm{km\ s^{-1}}$ before the satellite's infall and reaches its peak velocity (60~$\rm{km\ s^{-1}}$) at the first pericenter. 
    In the low-mass satellite analog ($m12i$) the outer halo does move with respect to the inner halo but only up to 20~$\rm{km\ s^{-1}}$. In the MW--LMC idealized simulations the magnitudes of both the $v_{com, outer}$ and $v_{host}$ are similar to those found in the MW--LMC analog ($m12b$).}
    \label{fig:vcom}
\end{figure*}

We start by studying the velocity of the host center of mass ($\mathrm{v_{COM}}$) as a function of 
time. This is shown in Figure~\ref{fig:vcom}, where the $\mathrm{v_{COM}}$ used is the one reported 
by the halo finder. The reference frame is the box, and as a result, all of the halos have a 
positive Lagrangian velocity as they travel through the cosmic web. For reference, the 
pericentric passages of the massive satellites are shown as vertical dashed (blue) lines. 
In the MW--LMC idealized simulations the reference frame is centered at the center of the simulated 
volume where the halo was placed at rest at $t=11.5$ Gyr. As the massive satellite passes through the pericenter,
the host's $\mathrm{v_{COM}}$ rapidly changes as the inner halo and disk are accelerated by the satellite.

In the MW--LMC analog ($m12b$), the host halo's velocity suddenly decreases by $\approx 80$ $\rm{km\ s^{-1}}$ during the first 
pericenter of the satellite (shaded regions of in the top right panel). At the second pericenter, the velocity increases by $80$ $\rm{km\ s^{-1}}$, 
illustrating the host halo reacting to the satellite. These results are consistent with those recently presented by \cite{Salomon23COMshifts}. Note that the time scales of these 
changes occur over $\approx$2~Gyr (time between pericenters, shown in grey in Figure~\ref{fig:vcom}) in the MW--LMC analog ($m12b$), but the first decreases in velocity occur
over $\approx$0.5~Gyr. A 0.5~Gyr time-scale is comparable to the dynamical time of the halo at 
50~kpc and hence orbits beyond 50~kpc will not react adiabatically to the COM motion.

In the MW--LMC idealized simulations, changes in the velocity of the host $v_{\mathrm{host}}$ can be up to $\approx$60~$\rm{km\ s^{-1}}$ at 
the pericenter (bottom right panel). An important difference is that the first increase happens over $\approx$2~Gyr,
which is 4~times longer than the case in $m12b$ where the velocity of the host drops within 0.5~Gyr. Such a difference could be due to the 
different eccentricities of the orbits, as shown in Figure~\ref{fig:orbits_properties}, 
and potentially to differences in the halo response between a cosmological and idealized halo. In our low-mass satellite analog ($m12i$) we see that both $\mathrm{v_{host}}$ and $v_{COM,outer}$ 
are roughly constant through the evolution of the galaxy, confirming that the satellites in $m12i$ 
are not massive enough to cause significant perturbations in the halo.

The fast motion induced by the massive satellites on the host halo is shown in 
Figure~\ref{fig:vcom} will induce a COM motion and \textit{reflex motion} between the inner and outer halo. As shown in GC21, this will be the 
main cause of apparent changes in orbital poles, which we discuss further in \S~\ref{sec:all-sky}. A 
full quantification of the reflex motion in the FIRE halos will be presented in Riley et al., in 
prep. Here we compute the reflex motion using the subhalos of the MW host, 
and excluding subhalos brought in with massive satellites. The lower panels in Figure~\ref{fig:vcom} 
show the relative velocity between the outer halo subhalos and the $\mathrm{v_{COM}}$ of the disk 
computed with the halo finder. The outer halo was arbitrarily defined to be the region beyond 50~kpc. 
We find that the reflex motion is stronger for the MW--LMC analog ($m12b$). As 
expected, the changes happen at the pericenter. In particular, the first pericentric passage 
induces the largest reflex motion $\approx$30~$\rm{km\ s^{-1}}$. 

We find that the amplitude of the reflex motion in the \latte{} halos is in the range of 30--80~km/s, similar to the reflex motion measured in the MW halo due to the LMC's infall ($\geq$ 34~km/s; \citealt{Petersen20}) . Among all the \latte{} halos the reflex motion experienced by the MW--LMC analog ($m12b$) is the closest to the MW--LMC, validating our choice of this system as the best analog. 

In the following section, we will explore how the perturbations caused by massive
satellites in the host halo -- including the reflex motion -- affect the distribution 
of orbital poles of star and dark matter particles, as well as bound substructures.

\subsection{Qualitative evolution of the distribution of orbital poles due to interactions 
            with massive satellites}\label{sec:all-sky}

\begin{figure*}[ht]
    \centering
    \textbf{\Large All-sky distribution of orbital poles in the outer halo} \par\medskip
    \textbf{\hspace{10mm} MW--LMC idealized sims.  \hspace{10mm} MW--LMC analog    (\textit{m12b}) \hspace{10mm} Low-mass satellite analog (\textit{m12i})}\\
    \textbf{Infall} \par\medskip
    \includegraphics[scale=0.26]{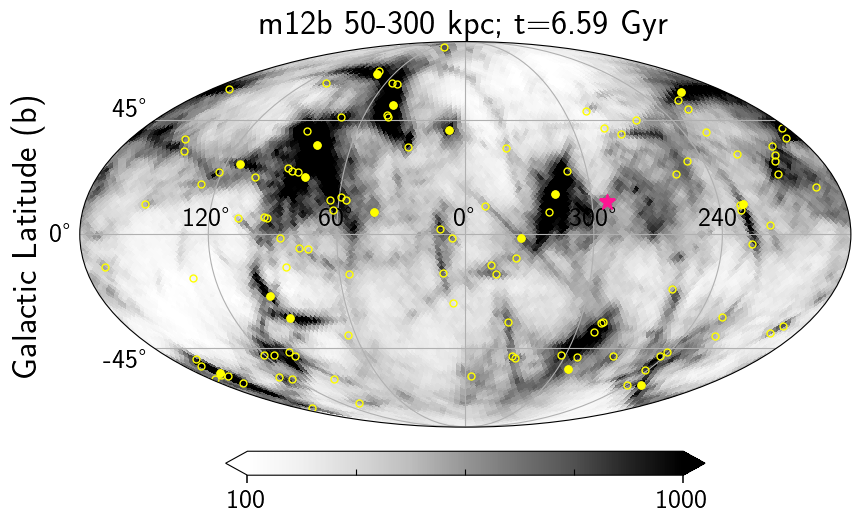}
    \includegraphics[scale=0.26]{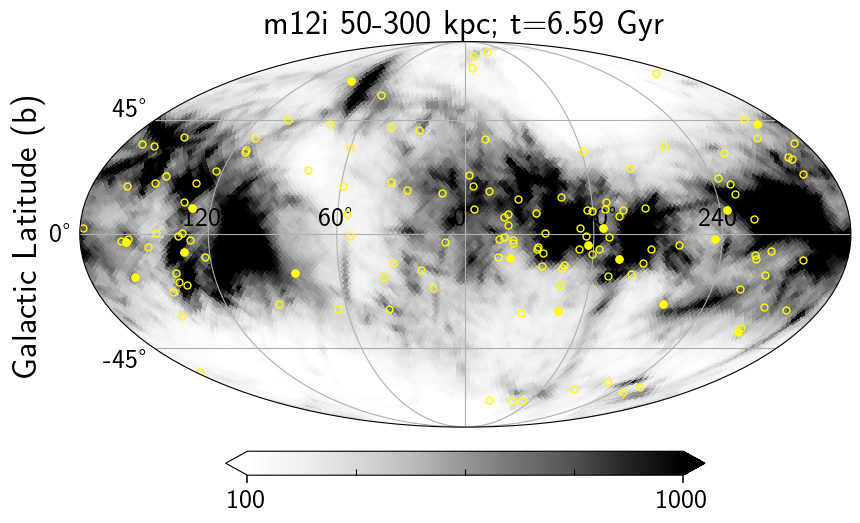}
    \includegraphics[scale=0.26]{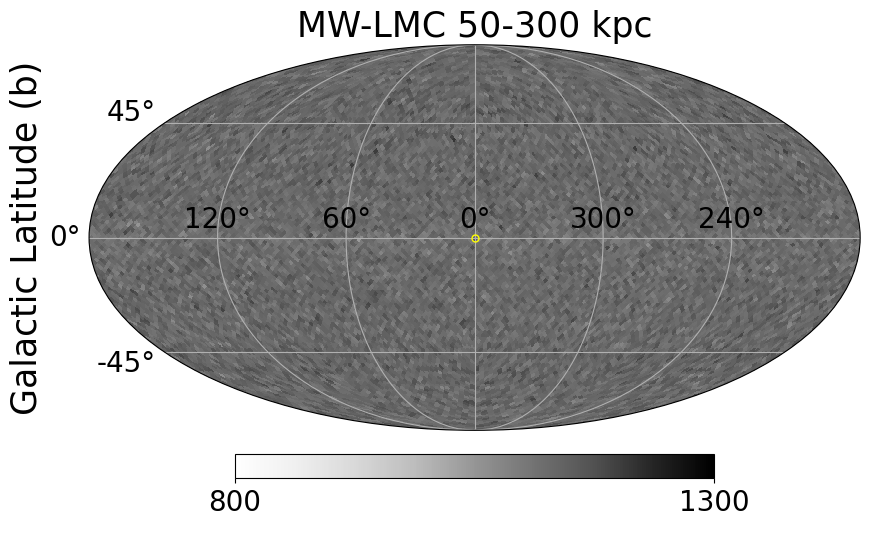}

    \textbf{1st pericenter} \par\medskip
    \includegraphics[scale=0.26]{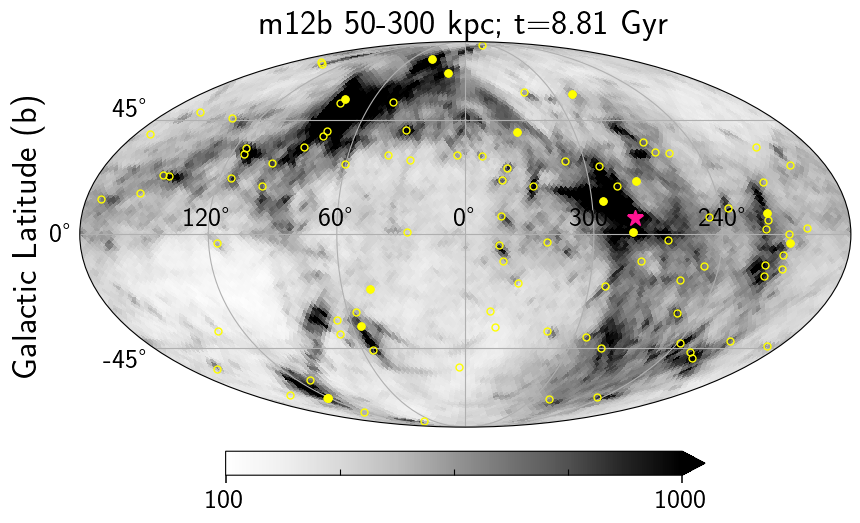}
    \includegraphics[scale=0.26]{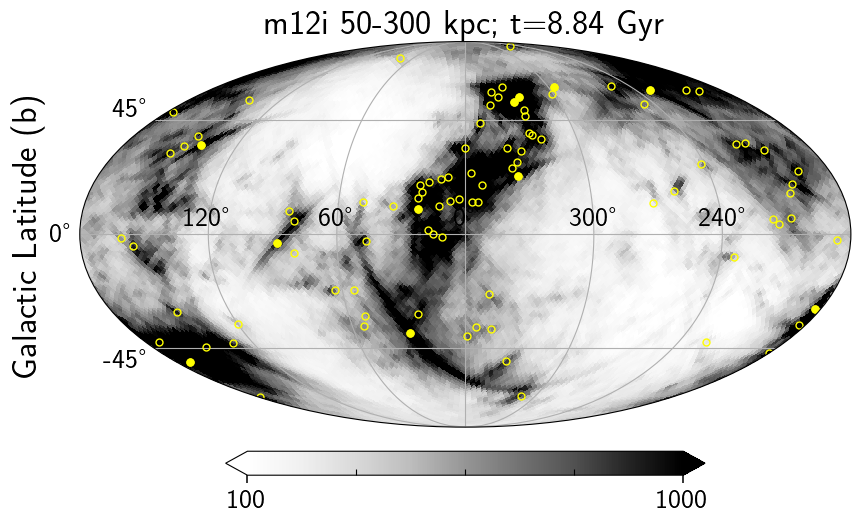}
    \includegraphics[scale=0.26]{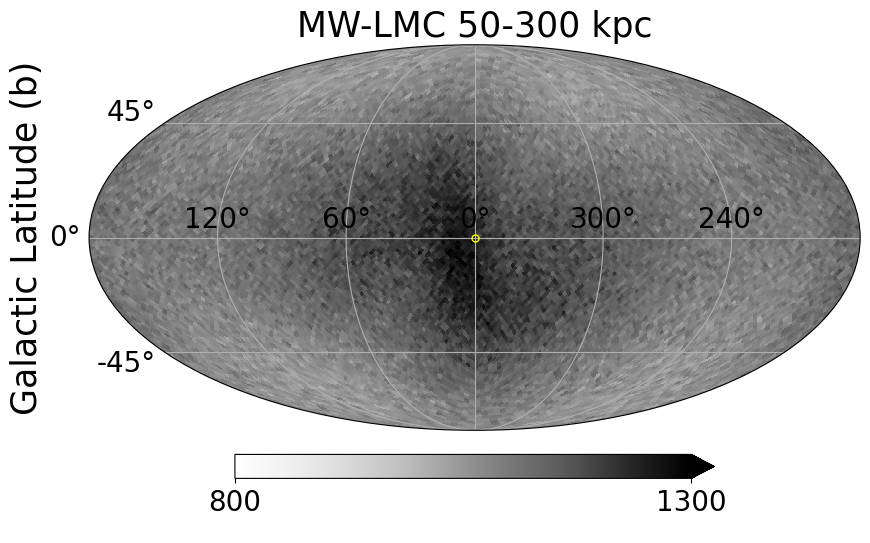}

    \textbf{2nd pericenter} \par\medskip
    \includegraphics[scale=0.26]{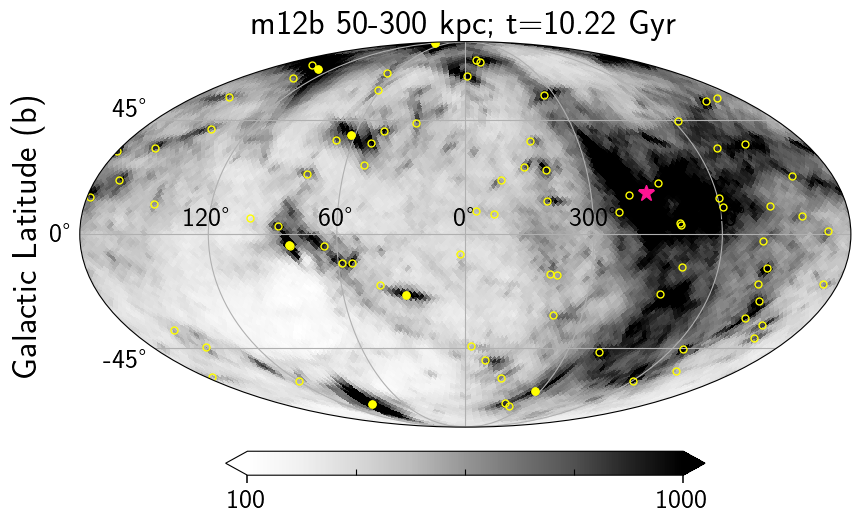}
    \includegraphics[scale=0.26]{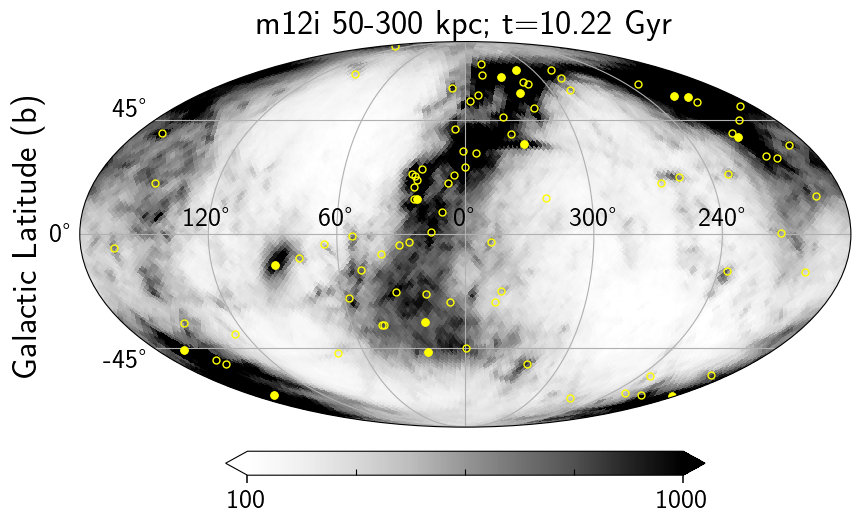}
    \includegraphics[scale=0.26]{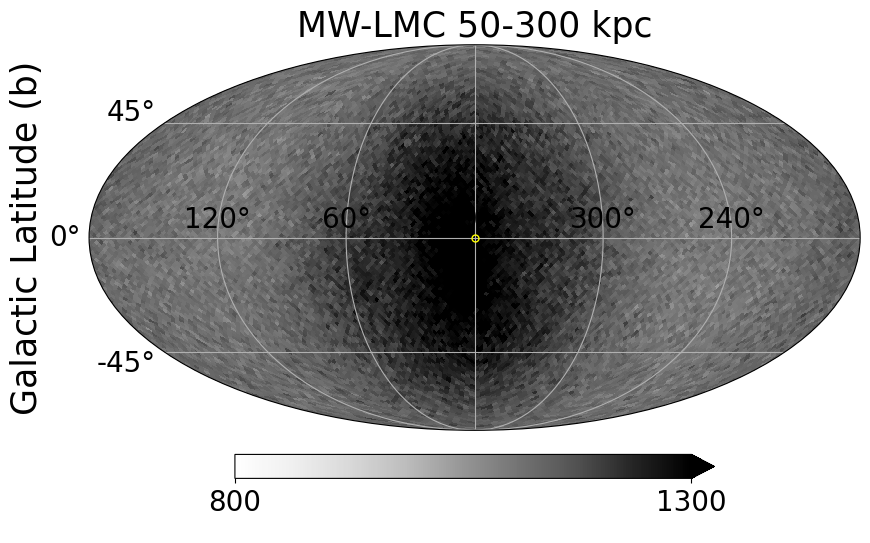}

    \caption{All-sky orbital poles density distribution for dark matter particles of the host (grey color map) between 
    50-300 kpc, note that the DM particles of the massive satellite are excluded in all cases. Orbital poles of subhalos and satellite galaxies are shown with open and solid yellow
    circles respectively. Each row corresponds to a different time in the evolution of the halo at the satellite's infall (top row), 1st pericentric passage (middle row), and 2nd pericentric passage (bottom row).   
    MW--LMC idealized simulation (right panels), MW--LMC analog $m12b$ (left panels), and the low-mass analog $m12i$ (middle panels). The orbital poles 
    of the particles were computed after rotating the disk into the x--y plane. The density of 
    orbital poles was computed using healpix with Healpy \citep{healpy}. In $m12i$ the orbital poles 
    distribution evolves smoothly from the overdensities of poles close to the equator 
    (upper middle panel) to a co-latitudinal distribution of poles (middle and bottom panel).
    In $m12b$ the distribution of orbital poles 
    changes rapidly after the first pericentric passage (middle left panel), and afterward, clustering of 
    poles is seen around the satellite's orbital pole  (pink marker) at $b=15^{\circ}, l\approx 270^{\circ}$). Similarly, in the MW--LMC idealized simulations the clustering occurs at the pericentric passage around the satellite's pole at $l=0^{\circ},\ b=0^{\circ}$}.
    \label{fig:all-sky}
\end{figure*}

As shown in the previous section, the outer halo appears to be moving in a 
galactocentric reference frame as the massive satellite orbits. 
Thus, the orbital poles of outer halo tracers are expected to show evidence of the
apparent motion of the outer halo.

Figure~\ref{fig:all-sky} shows the distribution of orbital poles of outer halo tracers
($\geq 50$~kpc\footnote{Note that a different cut in the outer halo between 30--60 
kpc does not considerably change the results}) for different host halos 
(shown in each column) and as a function of time (shown in each row). The orbital poles 
are from the tracers defined in Table~\ref{tab:subhalos_selection},  host DM particles (grey color maps), all subhalos (yellow empty circles), and satellite galaxies (solid yellow circles). In all cases, particles 
from the host's disk and the LMC-analog (provided by the halo finder at the time of peak mass of the satellite) were removed. 
As discussed in Section~\ref{sec:methods_ops}, we chose a Galactocentric reference frame where the disk lies on the $x-y$ plane. We also kept the longitude constant across snapshots to guarantee that the reference frame is the same and does not rotate between snapshots.
The top row of Figure~\ref{fig:all-sky} shows orbital poles at infall of the massive satellite, when the massive satellite in the MW--LMC analog ($m12b$)
is at the virial radius of the host. The middle row is close to the first pericentric passage of
the satellite and the bottom row is close to the second 
pericentric passage. In the low-mass satellite analog ($m12i$) we show the same snapshots as for the MW--LMC analog ($m12b$).

In the idealized MW--LMC simulations (right panels in Figure~\ref{fig:all-sky}) the initial distribution of orbital poles is isotropic (top left right).
However, after the first pericentric passage of the satellite, the orbital poles cluster around the orbital poles of the LMC at $l=0^{\circ}, b=0^{\circ}$ 
(middle panel) and this clustering lasts for the entire time of the simulation (8~Gyrs) (lower left panel). As discussed in \cite{GC21planes}, this clustering is 
\textit{apparent}, in the sense that it is sensitive to the reference frame of the observer. 
The dynamical times in the MW's halo change as a function of Galactocentric radius 
such that the inner regions of the halo respond faster than the outer parts to the LMC. 
As a result, there is a differential motion between the inner and outer halo. 
In a reference frame centered on the disk, the outer halo thus appears to be
co-rotating with the LMC and hence the orbital poles of the outer halo tracer appear clustered.

In a cosmological halo (left and middle panels of Figure~\ref{fig:all-sky}),
 the distribution of orbital poles is anisotropic at all times as a result of the 
 substructure and continuous of accretion into the host.
DM and star particles belonging to a subhalo have the same angular momentum as measured 
from the center of the host galaxy. Those stars and particles appear 
clustered in orbital pole space, as shown by the yellow circles (satellite orbital poles) overlapping
 with denser/darker regions of the DM orbital poles map. As substructure is disrupted by the tidal 
 forces from 
the host, particles will still move coherently (i.e., with the same angular momentum direction) and 
have clustered orbital poles, even though 
they are distributed along tidal tails (i.e., they do not cluster in position space). This idea has 
been applied to find substructure, such as streams, in the MW \citep[e.g.,][]{Johnston96, Mateu17}.

Accretion through filaments also leaves characteristic patterns 
in the distribution of orbital poles. Orbits of subhalos accreted from the same filament 
tend to be co-planar. A planar co-orbiting structure will be clustered in orbital poles, as the direction
of the angular momentum of the subhalos is the same. 
As the subhalos orbit the host, their angular momentum results in a sinusoidal distribution of orbital poles 
rather than just being clustered \citep[e.g.,][]{Lovell11}. 
In the low-mass analog ($m12i$, right panels of Figure~\ref{fig:all-sky}) the initial distribution of orbital 
poles tend to be aligned at galactic latitudes of $b=0^{\circ}$. 
Such anisotropic distribution is indicative that the accretion at that time was mainly along polar orbits 
(perpendicular to the host galaxy disk). This distribution changes 
gradually over time towards a sinusoidal global pattern where the poles are mainly aligned along galactic 
longitude $l=0^{\circ}$. The evolution of 
the orbital poles in the low-mass satellite simulation ($m12i$) is smooth on time scales of $\approx $2~Gyrs between $t=7-9$~Gyrs in the simulation. At this time there 
is a low-mass satellite orbiting $m12i$'s halo, which merges at 
$t\approx11$~Gyrs. This satellite is on an orbit with an eccentricity of $0.65$ 
(see Figure~\ref{fig:orbits_properties}). Since this is the most massive satellite in $m12i$ 
during that period of time, it is suggestive that the evolution of the orbital poles of the 
host halo DM particles is related to the passage of the satellite. 
After 9~Gyr there is no major evolution in the distribution of orbital poles in $m12i$.

In the MW--LMC analog ($m12b$, left panels of Figure~\ref{fig:all-sky}) on the other hand, 
the distribution of orbital poles is more isotropic at 6.59~Gyrs (left top panel). The isotropic distribution of poles is 
quantified in Figure~\ref{fig:delta_sph} where the spherical standard distance $\Delta_{sph}$ is lower in $m12i$ implying a higher 
isotropic distribution of poles. We also quantified that the initial distribution of poles in $m12b$ is more isotropic than in $m12i$ 
using the two-point correlation function (see \S~\ref{sec:twop_corr}, but we do not include these figures here. Presumably, the 
accretion in $m12b$ halo is more isotropic than in $m12i$ but a characterization remains to be done for these systems and it is outside the scope of this paper.

Regardless of the initial distribution of poles, what is important for us is the time 
evolution of the host poles. During the first pericentric passage of the satellite, the distribution of orbital poles experiences a rapid change\footnote{While we note that the timescale of the changes aren't visible from Figure~\ref{fig:all-sky}, followed by a second 
rapid change close to the second pericentric passage. Such rapid changes demonstrate that
massive satellites affect the apparent distribution of orbital poles mainly at pericenters. Furthermore, we see clustering around the satellite orbital poles ($l\approx 270^{\circ}$, $b\approx 15^{\circ}$) 
between the last two pericenters in this system. The existence of such clustering in a cosmological halo near LMC analog pericenters confirms the results obtained in the idealized simulations (right panels). We also note that at the second pericenter, there is not clustering of satellites, but mainly of subhalos and DM particles.}

To summarize, we find that in the
presence of a massive satellite, the distribution of orbital poles of tracers in the outer halo changes rapidly close to 
the pericentric passages of the massive satellite. The resulting distribution at pericenter is not representative of the 
distribution of orbital poles at the satellite's infall. 
In the absence of massive satellites ($m12i$), the distribution of orbital poles experiences a gradual evolution. But, it is worth
noting that in  $m12i$ and $m12m$ (see left panels in Figure 13 of the appendix) both galaxies absent of massive satellites, the distribution
of orbital poles at the start of our analysis $t\approx6$ Gyr is highly anisotropic, even more than in systems with massive satellite such as 
$m12b$. This highlights, that although massive satellites induce co-rotation and anisotropic distribution of orbital poles it is not the only 
mechanism. Most likely the accretion through filaments also plays a major role in the initial distribution of orbital poles. 
We will quantify these findings in the following section.

\subsection{Two-point correlation function analysis of orbital poles evolution}\label{sec:corrfunc}

\begin{figure*}[ht]
    \centering
     \Large{Correlation function relative to the snapshot at infall} \par\medskip
    \includegraphics[scale=0.6]{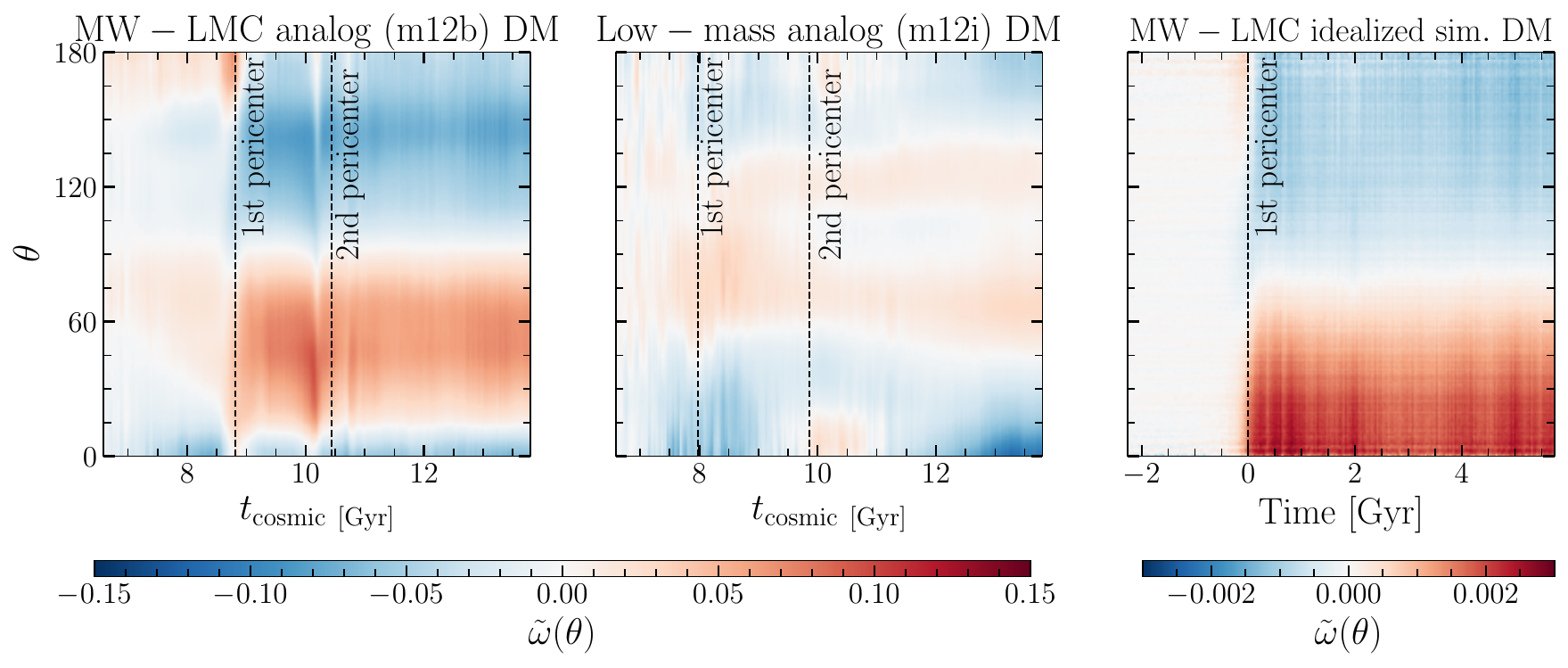}
    \raggedright     
    \includegraphics[scale=0.55]{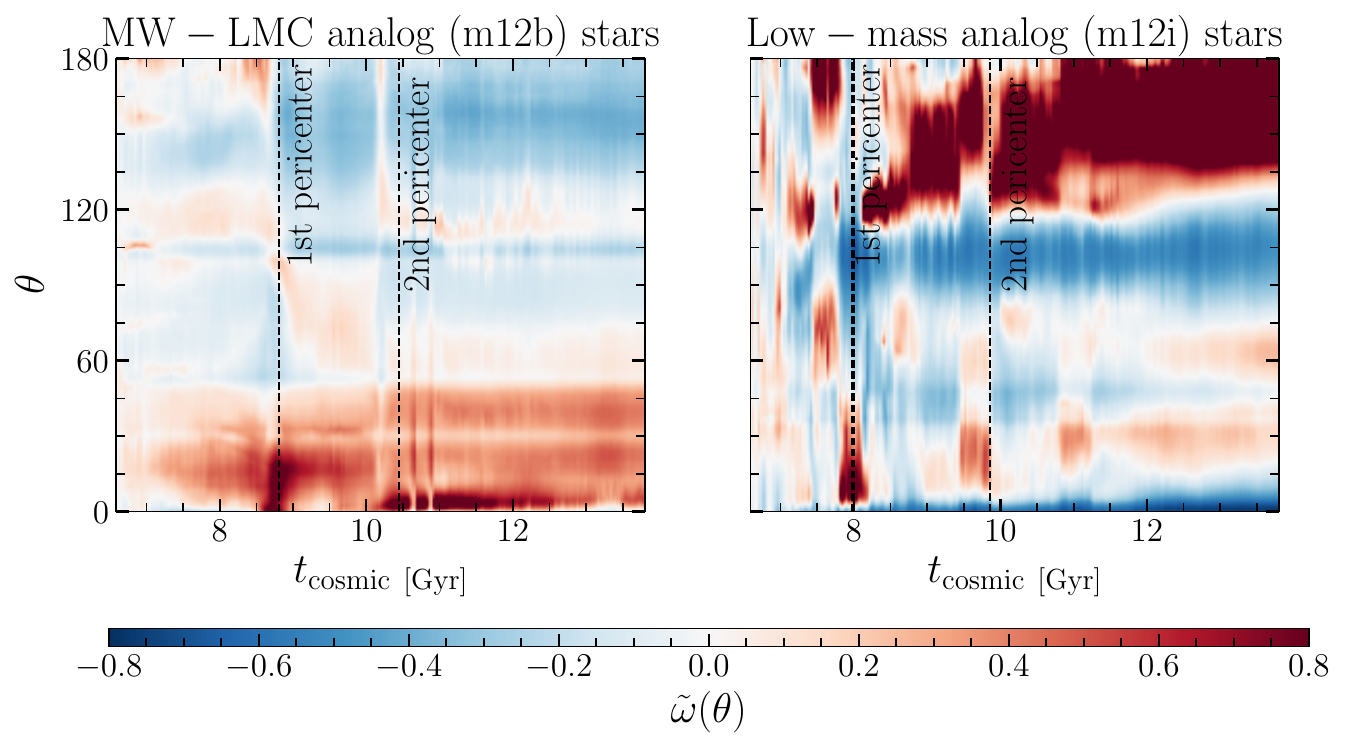}

    \caption{Temporal evolution of the two-point angular correlation function,
      $\tilde{\omega}(\theta)$, of the distribution of orbital poles for the MW--LMC analog ($m12b$), low-mass satellite ($m12i$), and idealized MW--LMC halos. In the top (bottom) panels the correlation function was computed on $10^6$ DM (star) particles randomly selected within 50-300 kpc. The distribution of orbital poles is roughly constant in time until a satellite approaches its pericenter (vertical dashed lines). Even minor mergers, such as the
one experienced by $m12i$ at $t=8$ Gyrs, change the orbital poles
distribution. Note that for the idealized MW--LMC halo (top right panel) the relative changes are two
orders of magnitude lower than both $m12i$ and $m12b$ (see text for further discussion on this). The correlation function computed in the star particles (bottom panels) varies more due to the presence of large stellar streams and substructures. Changes close to pericentric passages are present, but there are also many more changes that occur through the evolution of the halos. }
    \label{fig:2d_corrfunc_alldm}
\end{figure*}

\begin{figure*}[ht]
    \centering
    \includegraphics[scale=0.45]{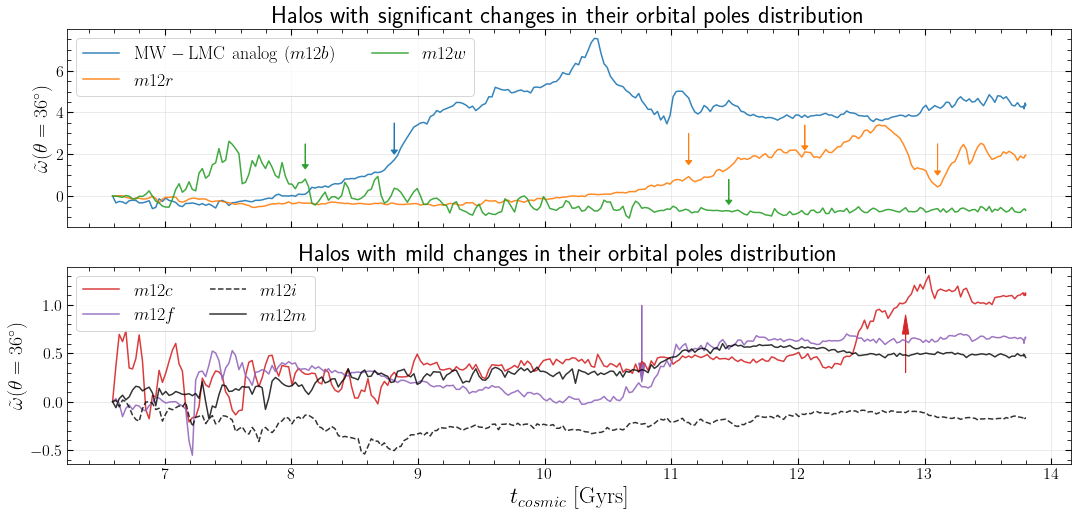}
    \caption{Summary of the temporal evolution of the relative correlation function $\tilde{\omega} (\theta)$ at  $\theta=36^{\circ}$ (same angular scale of the orbital pole dispersion of the 7 more clustered satellites in the VPOS). As in Figure~\ref{fig:2d_corrfunc_alldm} the correlation function was computed in $10^6$ randomly chosen DM particles in the outer halo (50-300~kpc). Positive (negative) values of $\tilde{\omega} (\theta)$ indicate an enhancement (decrease) of clustering within $\theta=36^{\circ}$.
    The 5 \latte\ halos with massive satellites are shown with color lines while the two halos without massive satellites are shown in black. The arrows represent the first pericentric passage of a massive satellite in each halo. In most of the halos, changes in $\tilde{\omega} (\theta)$ occur after the pericentric passages. The strongest clustering at this angular scale is seen in $m12b$ (blue line), which is the best LMC analog. However, there is also a noticeable enhancement in clustering in $m12r$ (orange line) which has three massive satellites. In systems that do not have a massive satellite during this time-window, like $m12i$ and $m12m$, 
    these changes in $\tilde{\omega} (\theta)$ are within 0.5 which is the lowest across all the halos}.
    \label{fig:summary}
\end{figure*}

In this section, we quantify the distribution and evolution of orbital poles of tracers in the host galaxy
 using the two-point correlation function introduced in 
Section~\ref{sec:twop_corr}. The angular two-point correlation function $\omega(\theta)$ measures the probability of finding a pair of orbital poles at an 
angular separation of $\theta$. For example, if the orbital poles are clustered within 30$^{\circ}$ the probability of 
finding pairs of poles at an angular scale smaller than $30^{\circ}$ would be greater than
at scales larger than $30^{\circ}$. As a result $\omega(\leq 30^{\circ}) \geq \omega(\geq 30^{\circ})$; see 
Equation~\ref{eq:corrfunc}. If the enhancement of poles is along a great circle in the sky --- as the one shown in 
Figure~\ref{fig:all-sky} for $m12i$ (right column, middle panel) --- the probability of finding pairs at scales near $0^{\circ}$ and at larger scale corresponding to the radius of the great circle close to $180^{\circ}$ would be large. 

With this intuition in mind, we compute $\tilde{\omega}(\theta)$ (the clustering relative to that at time $t_\mathrm{infall}$) as
a function of time for all the host DM particles (upper panels) and star particles (bottom panels)
within 50--300 kpc. We randomly select $10^6$ DM and star particles (excluding those of the massive satellite) in the outer halo (between 50--300 kpc). This is shown 
in Figure~\ref{fig:2d_corrfunc_alldm}, for the MW--LMC analog ($m12b$) 
(left panel), the low-mass satellite ($m12i$) (middle panels), and the MW--LMC halo (top right panel). Red (blue) colors show regions with higher (lower) 
relative probability of finding pairs of orbital poles, with respect to the snapshot at infall of the massive satellite in $m12b$ ($t=6.8$~Gyr for $m12b$ and $m12i$).
For the idealized MW--LMC simulation, we use the snapshot at $t=-2$~Gyr to compute $DD_{\textrm{infall}}$ (see \S~\ref{sec:twop_corr} for definitions), which is when the 
massive satellite was at the virial radius in the idealized simulation. At that time, the halo was unperturbed and hence the 
distribution of poles was isotropic, and $\tilde{\omega} (\theta)$ was zero across all angular scales. 

For the DM particles (upper panels) of MW--LMC analog ($m12b$) and the MW--LMC idealized simulation it is clear that after pericenter (vertical dotted black lines in Figure~\ref{fig:2d_corrfunc_alldm}) the distribution of 
poles changes drastically. In the MW--LMC analog ($m12b$) there is an enhancement in the probability of finding pairs at scales $\leq 
90^{\circ}$ and a decrease at scales $\geq 90^{\circ}$. This confirms that after pericenter the orbital poles 
distribution becomes less isotropic and more clustered. This can also be seen in the orbital poles all-sky movies of 
Table~\ref{fig:all-sky}. A very similar pattern is seen in the idealized MW--LMC case, where the clustering is enhanced at scales 
$\leq 60^{\circ}$ after the first LMC's pericenter. 

After the second pericentric passage, there is a rapid change in the distribution of orbital poles around $t=10$~Gyr, 
but this perturbation does not induce a long-term evolution as the first 
pericentric passage does. The effect of the second pericentric passage is not seen as strongly in the idealized MW--LMC simulations. 
Another difference between the \latte{} simulations ($m12b$ and $m12i$) and the idealized simulation is the amplitude of $\tilde{\omega}(\theta)$, which is two orders of 
magnitude stronger in the \latte{} halos. It is not clear why this is the case, because as shown in \citet{Weinberg22}, a 
Hernquist halo (used in the idealized MW--LMC simulations) is more stable than an NFW/Einasto halo. We also assume a spherical system fully in equilibrium, which is not the case in a cosmological simulation where halo shapes are constantly being affected by filaments. 
Such a difference in the initial dynamical state of the halos could be responsible for the difference in
the amplitudes of perturbations in the cosmological versus isolated halos.

In the halo interacting with the lower-mass satellite, $m12i$ (middle panel in Figure~\ref{fig:2d_corrfunc_alldm}), 
the amplitude of the changes in $\tilde{\omega}
(\theta)$ is lower than in the MW--LMC analog ($m12b$) halo. 
However, the changes are also correlated with the pericentric passages of the lower-mass satellite,  
mainly between 7.5 and 11 Gyrs. Note that in the MW--LMC analog ($m12b$) and in the idealized MW--LMC simulation, the change 
happens abruptly right after the satellite's pericenter at 8.6~Gyrs. Even though there is a lower-mass satellite in $m12i$, we 
can see weak changes in the distribution of orbital poles while the satellite orbits the host around 6--9~Gyrs. This confirms 
that even lower-mass satellites (like Sagittarius in the MW) can induce changes in the distribution of orbital poles in their host halos.    

Similarly, we randomly select star particles (excluding those of the massive satellite) in the
 halos of both the MW--LMC analog ($m12b$) and in the low-mass satellite analog ($m12i$). In the 
 stellar halo (lower panels of Figures~\ref{fig:2d_corrfunc_alldm}), we observe changes in the 
 correlation
function at similar times as those observed in DM particles. Yet, there is more noise in 
the signal, most likely 
due to the substructure in the stellar halo.  In $m12i$ there is a large enhancement in the
 correlation starting at $t=8$~Gyrs. This is likely due to a large stellar stream produced 
 from the merger with a low-mass satellite that was not removed from our analysis.  Note also
  that the amplitude of the effects is larger in the stellar halo 
than in the DM halo. This is because the relative overdensities produced by substructures
 such as streams are larger in the stellar halo than in the DM halo. This suggests that 
 there should also be a signal of orbital-pole clustering in the MW stellar halo once 
 we have full 6D information for outer halo stars in the MW.

Another way to look at the changes in orbital-pole clustering is to compare the time evolution of 
power in the correlation function at a fixed angular scale for each of the simulations. 
Figure~\ref{fig:summary} shows the temporal evolution of $\tilde{\omega}
(\theta=36^{\circ})$, which roughly 
corresponds to $\Delta_{sph}(k=7)$ for the measured orbital poles of the VPOS \citep{Pawlowski21}, for all 
the \latte{} halos. A larger positive value of $\tilde{\omega}
(\theta)$ implies a larger likelihood of finding pairs at 
this angular scale 
($\theta=36^{\circ}$), or in other words clustering enhancement. 

For the seven \latte{} halos there are a total of ten massive satellites whose time of pericentric passages
are shown with the colored arrows in Figure~\ref{fig:summary}. In halos, $m12i$ (black solid
line) and $m12m$ (black dashed line) $\tilde{\omega}(\theta)$ do not change 
considerably ($|\tilde{\omega}(\theta)| \leq 0.5$) over time since they are not accreting massive 
satellites. In the halos that 
undergo mergers with massive satellites ($m12b, m12c, m12f, m12r, m12w$), temporal changes in $\tilde{\omega}(\theta)$ seem to be correlated with the pericentric passage of the satellite (shown with the arrows). However, we see a wide variety of amplitudes in 
$\tilde{\omega}(\theta)$. In some cases, satellites do not cause any change (second satellites in $m12w$ and in $m12r$) since their pericentric passages are either ($\leq 10$ kpc) or they are not massive enough to induce a significant change. In one case (first satellite passage in $m12w$ at $8.1\rm{Gyr}$), $\tilde{\omega}(\theta)$ can decrease, not that this is a head-on merger of a satellite with a high eccentricity. In other cases, the relative changes in $\tilde{\omega}(\theta)$  start before the pericentric passage like in $m12c$. These highlights, the variety of effects that satellites impart in the orbital poles distributions. These results suggest that satellites whose pericentric passage is within 10--60~kpc are the ones that induce 
more clustering. 

Out of all the halos, $m12b$ (blue line) shows the strongest change, illustrating that an LMC-like satellite 
induces the strongest enhancement in the clustering of orbital poles of the host. The variety in the orbital 
poles changes illustrates that satellites in orbits similar to the LMC and with similar masses have the strongest
effect on the hosts.

\section{Discussion}\label{sec:discussion}

\subsection{Effect of massive satellites on the dispersion of satellite and subhalos orbital poles}\label{sec:global}

\begin{figure*}[ht]
    \centering
    \includegraphics[scale=0.6]{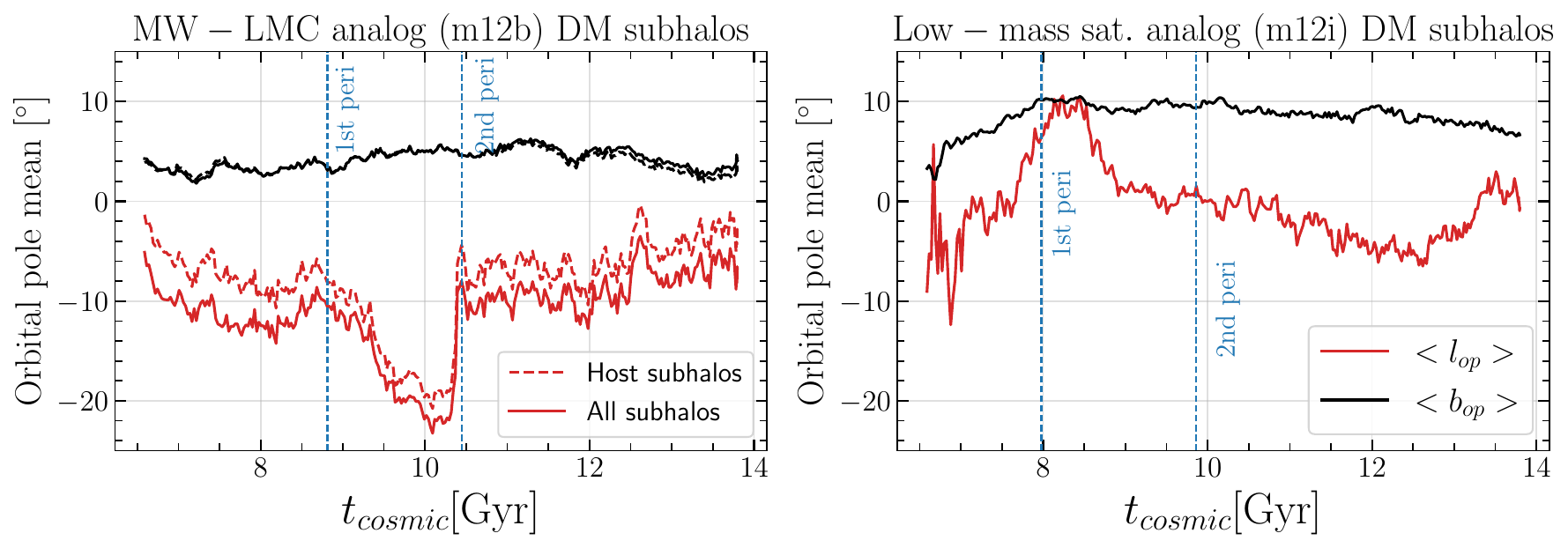}
    \includegraphics[scale=0.6]{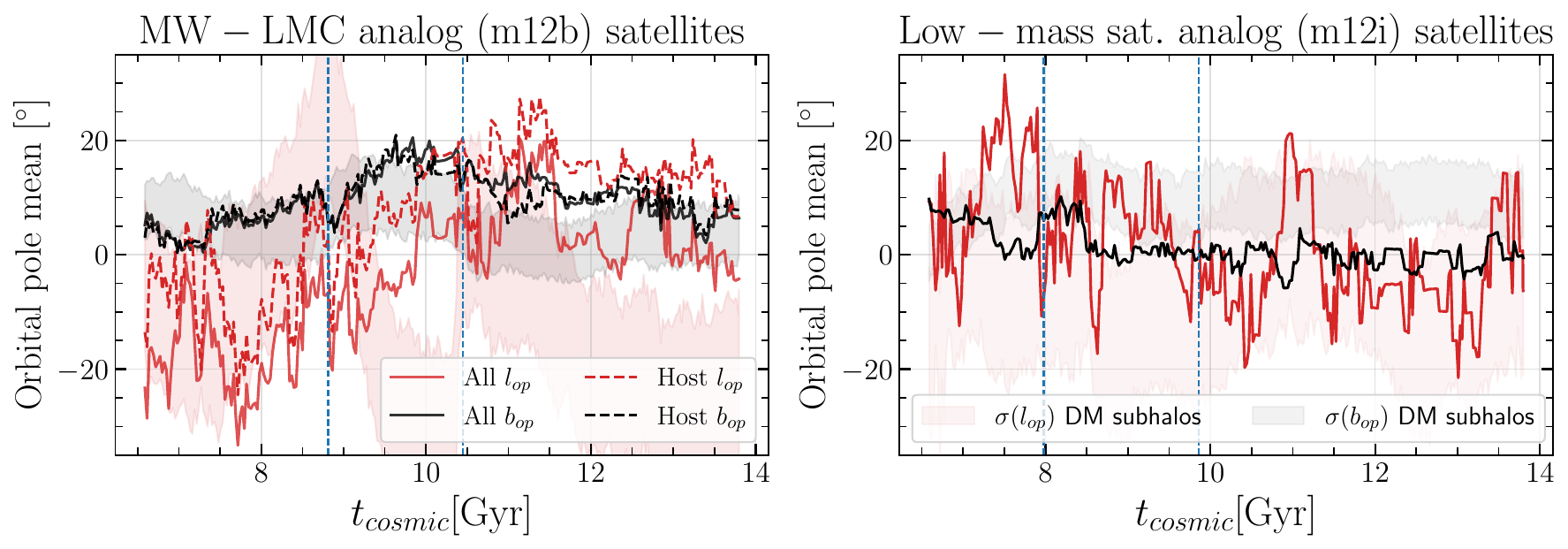}
    \caption{Mean values of the orbital poles of all the subhalos (upper panel) and satellite galaxies (bottom panel) with masses 
    $\geq 10^{7}M_{\odot}$ and within 50-300 kpc in $m12b$ (left panel) and $m12i$ (right panel). Variations in the mean 
    orbital poles are seen at the pericentric passages (vertical dashed lines) of the massive satellite galaxies. In both $m12b$ and 
    $m12i$ the changes are primarily in the longitudinal orbital poles $l_{op}$, after the pericenter of the satellites (blue 
    dashed lines). In $m12i$, there is an increase of the mean close to the passage of the satellite at $\approx 8$ Gyr. In 
    the satellite galaxies (bottom panels) the mean values are more noisy given the lower number of satellites compared to 
    subhalos. Shaded regions show the distribution of the mean orbital poles of DM subhalos after down-sampling to the same number of satellite galaxies. In $m12b$, the mean values of $l_{op}$ increase after pericenter the opposite trend as seen in the subhalos.}
    \label{fig:OP_mean}
\end{figure*}

In the previous section, we showed that the all-sky distribution of orbital poles evolves 
through the evolution of a halo. Massive satellites with pericentric passages 
between 10--60 kpc change the distribution of poles of all the halo tracers (satellites, subhalos, stars, and DM particles) in the host. We quantify these changes using the two-point 
angular correlation function. In this section, we compare our results to standard metrics used to quantify the co-rotation (clustering) of 
orbital poles in the context of satellite planes. We focus on the temporal evolution of the 
mean and dispersion of orbital poles.  

\begin{figure*}[ht]
    \centering
    \includegraphics[scale=0.6]{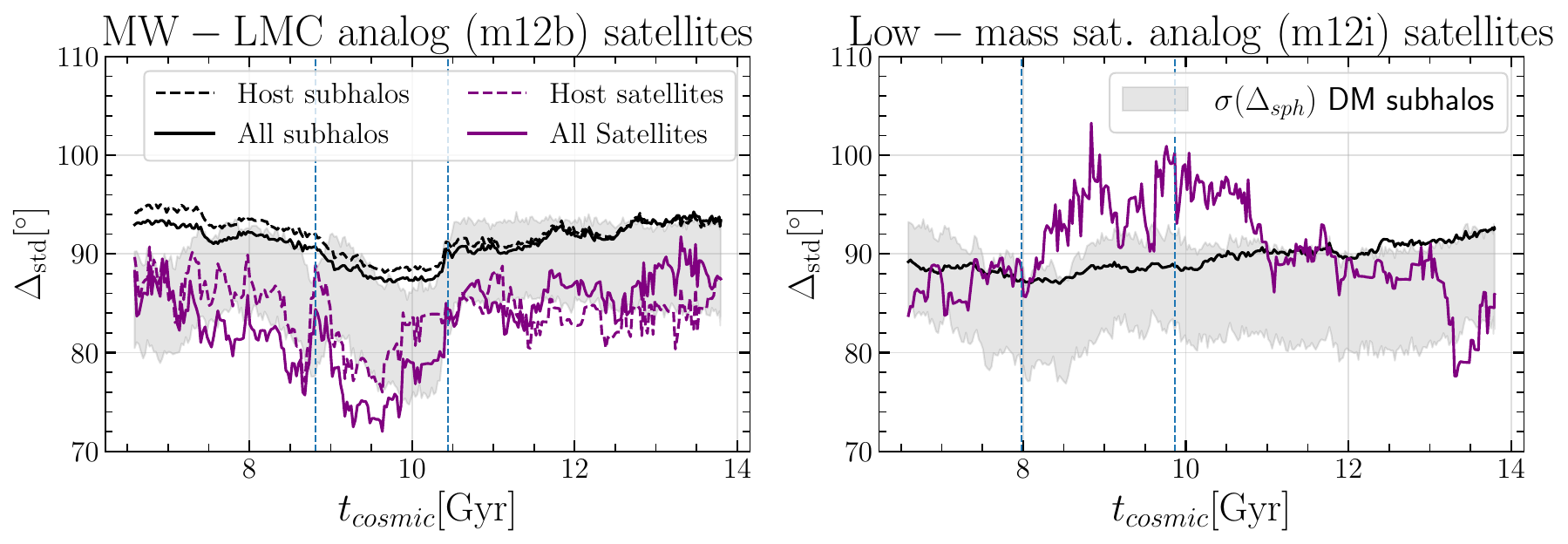}
    \caption{Dispersion of the orbital poles computed using subhalos with masses 
    $\geq 10^{7}M_{\odot}$ and within 300 kpc in $m12b$ (left panel) and $m12i$ (middle panel). 
    The grey-shaded regions show the dispersion in the orbital poles of the DM subhalos after 
    down-sampling to the same number of satellite galaxies. The dispersion in the orbital poles 
    of the MW--LMC analog ($m12b$) decreases for both subhalos and satellites between the pericentric passages of the satellites (blues dotted lines). Note that the decrease in the dispersion of the satellites is larger than in the subhalos. In $m12i$, on the other hand, the subhalos dispersion is unaltered, while in the satellites there is an increase after the pericenter of the low-mass satellite.}
    \label{fig:OP_disp}
\end{figure*}

\begin{figure*}[ht]
    \centering
    \includegraphics[scale=0.63]{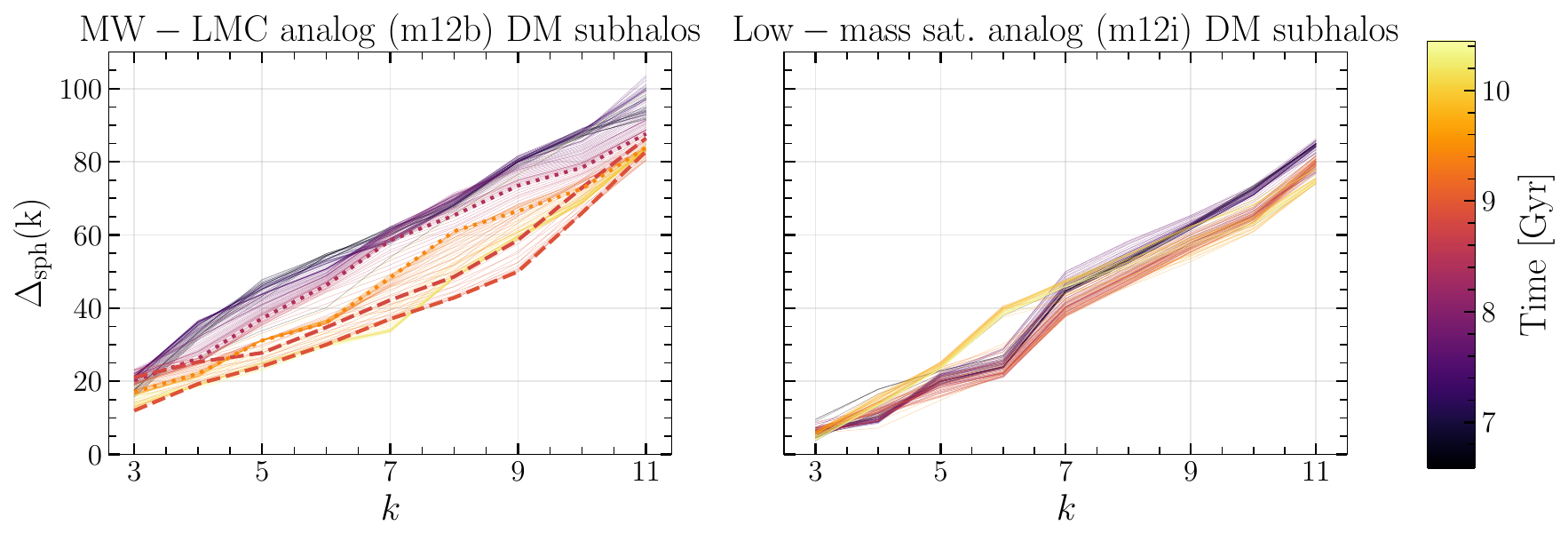}
    \caption{Temporal evolution of the spherical standard distance $\Delta_{sph}$ of the $k$ closest orbital poles of the top 11 massive satellites in the MW--LMC analog $m12b$ (left panel) and $m12i$ (right panel). Smaller values of $\Delta_{sph}$ correspond to more clustered orbital poles. During the first pericentric passage of the massive satellite in $m12b$, $\Delta_{sph}$ changes rapidly and reaches its minimum. Dashed lines show $\Delta_{sph}$ close to the pericenter passage, while the dotted lines show the times closer to the satellite's infall. In contrast with the evolution of $\Delta_{sph}$ in $m12i$, the spread in the values of $\Delta_{sph}$ reaches $\approx 30^{\circ}$, almost twice as in $m12i$.} 
    \label{fig:delta_sph}
\end{figure*}

We start by computing the mean of the orbital poles as a function of time for all the subhalos
with masses above $10^7~M_{\odot}$, in 
both the MW--LMC analog ($m12b$) and in the low--mass analog ($m12i$). This is shown in 
Figure~\ref{fig:OP_mean} for all of the DM subhalos (upper panels) and satellites (bottom panel) 
with masses $\geq 10^{9}\mathrm{M}_{\odot}$ in the MW--LMC analog ($m12b$) and in the low--mass analog $m12i$, upper left and right panels.
In both the MW--LMC analog ($m12b$) and low-mass satellite ($m12i$), the mean longitudinal component of the orbital poles 
 (red lines in the upper lines) change during the
pericentric passage of the satellites (dotted blue lines). This can be understood by looking at Figure~\ref{fig:all-sky} where 
the distribution of poles is symmetric in orbital poles latitude, but not in longitude. Note that this is not the case in all the \latte{} halos, as Figure~\ref{fig:all_sky2} shows that in $m12f$ and $m12m$ changes mainly happen in the latitudinal component of the orbital poles. In the MW--LMC analog ($m12b$), we also compute the mean orbital pole 
removing the satellites and subhalos of the massive satellite (dashed lines), by removing the subhalos of the satellite identified at the
satellite peak mass before infalling in the host. The mean of the orbital poles is always lower when all the subhalos are 
included, specially in the longitudinal component (red lines). This is intuitive since the massive satellite brings several galaxies in particular regions on the sky. Yet, when 
measuring the mean with only the host subhalos the decrease is $\geq10^{\circ}$ 
between pericenters (dashed lines in the top left of Figure~\ref{fig:OP_mean}.

For the satellite galaxies, the mean of the satellites' orbital poles (bottom panels in Figure~\ref{fig:OP_mean}) does 
not show the same behavior as the DM subhalos in MW--LMC analog ($m12b$). To understand this behavior, 
we down-sample the DM subhalos (shaded regions in the bottom panels) to compare directly
with the satellite population. The longitudinal component of the mean orbital pole in the MW--LMC analog ($m12b$) (left panel)
increases between pericenters, which is the opposite behavior of the subhalos. This is because the satellites are not 
randomly distributed among the subhalos, but rather biased to be around the massive satellites. We checked that 
when selecting only DM subhalos with peak masses $\geq 10^{8.5}$M$_{\odot}$ we achieved results consistent with the satellite galaxies.

Figure~\ref{fig:OP_disp} shows the temporal evolution of the dispersion of the orbital poles 
for all DM subhalos (black) and satellite galaxies (purple) with halo masses $\geq 10^{9}M_{\odot}$ computed with Equation~\ref{eq:op_disp}. 
Lower values in the dispersion indicate the clustering of orbital poles. 
The shaded region corresponds to the 68\% percentile after sampling the subhalos population with the same number of 
satellite galaxies. In $m12b$, $\Delta_{orb}$ decreases
by $\approx 10^{\circ}$ ($\approx 20^{\circ}$ for the satellites) at the massive satellite's pericenter 
(dashed vertical blue line). In $m12i$, on the other hand, the satellites (purple lines) exhibit 
a rapid increase of $\approx 10^{\circ}$ in the dispersion after the low-mass satellite's pericenter. Overall the effects 
are milder when measuring the dispersion in the $m12i$ population of subhalos (black lines).  Yet, our main findings are the same, changes in 
the mean of the orbital poles take place between the satellite pericenters. However, it is important to notice that the amplitudes in both the mean and dispersion are sensitive to the tracers population used. Satellite galaxies do not have the same mean as DM subhalos even when down-sampling their numbers to match that of the satellites. 

Similarly, we compute the spherical standard distance as defined in Equation~\ref{eq:op_sph}. For this, we select the 11 most massive satellites in the MW--LMC analog ($m12b$) and low-mass halo ($m12i$) over a period of time of $\approx$5~Gyr. We select those satellites that at $z=0$ are within the virial radius of the host. In the case
of the MW--LMC analog ($m12b$), this time window spans over the interaction with the massive satellite. The results are shown in 
Figure~\ref{fig:delta_sph}. We exclude subhalos of the massive satellite.

We found that in the MW--LMC analog ($m12b$) $\Delta_{sph}$ evolves widely and exhibits a larger range of values, mainly between 8--9 Gyrs, which is when 
the massive satellite makes it's first pericentric passage. For a given value of $k$, $\Delta_{sph}$ decreases by up to $30^{\circ}$ around when the massive satellite passes through first pericenter. In contrast, in low-mass satellite ($m12i$) the values of $\Delta_{sph}$ are more constant in time. For a given value of $k$, $\Delta_{sph}$ varies within 
$\approx$15$^{\circ}$. We also note that in the low-mass satellite halo analog ($m12i$) the orbital poles are more clustered (overall lower values of 
$\Delta_{sph}$) to begin with than in the MW--LMC analog ($m12b$). Note that compared with expectations from other cosmological simulations (e.g., see Figure~5 in \cite{Pham22}) these values are well within the expected ranges. However, we remind the reader that the initial 
distribution of poles is set by the accretion of the subhalos along the filaments, which is different in the MW--LMC analog ($m12b$) 
and the low-mass satellite halo ($m12i$). Here we have shown that even if the initial clustering was higher -- lower values of
$\Delta_{sph}$ -- in the low-mass satellite halo ($m12i$) the interaction with the massive satellite in $m12b$ enhances the clustering at 
pericenter where it becomes comparable to the clustering in the low-mass satellite ($m12i$).

\subsection{Satellite kinematics as tracers of the dark matter halo response}

The kinematics of satellite galaxies provide insights into the assembly history of the host galaxy, its total mass, and the shape of its DM halo. The results presented here, suggest that the orbital poles space can provide further insight into the dynamical state of the galaxy. The unusual co-rotation pattern seen in the satellites could be the results of several mechanisms, in particular, from filamentary structure \citep[][]{Libeskind11, Lovell11}, group infall \citep[e.g.,][]{Li08, Donghia08, Smith16, Vasiliev23LMC2}, from tidal dwarfs \citep{Hammer13, Wang20, Banik22}, mergers \citep{Smith16}, and the MW's halo response to the LMC passage. 
As discussed in \S~\ref{sec:kinematic}, the halo response is composed of several processes, but among all, the main one is the dipolar instability induced by the LMC. 

The dipolar mode with $l=1$ is the easiest to excite, and it is a \textit{weakly} damped mode, whose lifetime spans to times comparable to the age of the universe \citep{Weinberg22}.  key signature of a rotating dipole mode is the co-rotation in the outer halo as shown in this paper and in GC21b. 
For example, one can use the observational measurement of the VPOS to ask: what would be the necessary halo 
response to induce the observed co-rotation? As such, this is a new avenue to use the disequilibrium state of the galaxy to constrain the DM halo response in the out-of-equilibrium regime.

Other possible observational signatures of the DM halo response could potentially be seen in the lopsidedness of stellar disks as proposed in \citep{Varela-Lavin23}.

\subsection{Placing the kinematic state of the MW in a cosmological context}

Placing the MW in a cosmological context is key to interpreting the MW's formation. Large box cosmological 
simulations are very useful given the large number of MW-like galaxies in these simulations that can be 
used to statically study the formation of the MW. Indeed, the planes of satellites is a problem given 
their rareness in MW-like galaxies in large box cosmological simulations. The MW is at a particular point in its 
evolution given that the LMC just passed its first pericentric passage. As such, comparing the dynamical state 
of the MW should be done with MW-like galaxies that are in similar dynamical states. Here we discuss key 
conditions that need to be taken into account when comparing the MW with cosmological simulations in the 
context of the planes of satellites.

As shown in the previous sections, the presence of massive satellites near the pericenter coincides with enhanced clustering of orbital poles in the halo DM particles, subhalos, and satellites of MW analogs. We found that the effect is maximal at the first pericentric passage of the 
massive satellite. As such, MW-like galaxies with massive satellites at their pericenter (at present-day) are ideal to 
compare with the MW. If one studies generic satellite mergers that occur at earlier times in the evolution of the host, the effects from the satellite might be negligible \citep[e.g.,][]{Kanehisa23}. 
The amplitude and change in the orbital poles 
distribution of the host depend on the orbit and host--satellite mass ratios as shown in 
Figure~\ref{fig:2d_corrfunc_alldm}. A low host--satellite mass ratio 
will not displace the inner halo COM significantly, but a massive satellite will. Very eccentric orbits will not induce apparent co-rotation, but rather a linear displacement of the inner halo. In contrast, more tangential orbits would induce 
apparent circular motion of the inner halo. Lastly, we found that the clustering of poles is higher for satellites whose pericentric distances are between 10--60~kpc. This is consistent with what is observed, since the 
LMC's pericenter distance is $\approx$50~kpc. 

Another important consideration for placing the MW in a cosmological context is the role of filamentary accretion.
As discussed in many previous works \citep[e.g.,][]{Libeskind11, Lovell11}, the location of the filaments set the distribution of orbital poles
in a galaxy. For the seven halos that we studied the distribution of orbital poles is distinct in all of them given that 
all of them reside in different filamentary structures. A natural question to ask then is, what has to be the orbit 
of the LMC relative to the direction of accretion of the other satellites to create the observed co-rotation planes of 
satellites? This would provide a path forward constraining the direction of accretion into the MW.

\section{Conclusions}\label{sec:conclusions}

We have characterized the apparent evolution of orbital poles of subhalos and satellite galaxies
in simulated galaxies that are undergoing LMC-like mergers. We used the \latte{} suite of FIRE-2 cosmological zoom-in simulations, where 
we focus on the last evolutionary stages of the halos, from $z\approx1$ to $z=0$. In the main text, we
present the results from two representative halos, a MW--LMC analog ($m12b$) with a satellite halo mass 
ratio of 1:5 merger, and a system merging with a low-mass satellite ($m12i$) representing a 1:12 merger. We also included in our analysis the idealized N-body 
simulations of \cite{garavito-camargo19a}, but run further in time for a total of 8~Gyrs. In the 
Appendix \S~\ref{sec:appendix} we expand our analysis to five more halos from the \latte{} suite.

We found that the distribution of orbital poles of the host (DM particles and stars)
change in the presence of satellite galaxies. 
Changes are stronger for massive satellites in particular after their first pericentric passages. 
Low-mass satellites like the one in $m12i$ also induce changes in the poles distribution but 
on longer time scales. The two-point angular correlation function allows to quantify changes at different angular 
distances $\theta$. We found that the changes in the orbital poles can happen at different
angular scales, suggesting that different satellite orbits induce changes 
at different scales. This is clearly seen in Figure~\ref{fig:2d_corrfunc_latte} in 
the appendix where we present the angular correlation function of all 
the \latte{} halos. Interestingly, the MW--LMC analog in $m12b$ produces the highest clustering at 
$\theta=36^{\circ}$, which is the scale of the observed clustering in the MW satellites \citep{Pawlowski12}.

Although we found that the presence of massive satellites does change the distribution of orbital poles in the five halos that we analyzed this is not the only mechanism. Halos without massive satellites such as $m12i$ and $m12m$ also show a highly anisotropic distribution of poles. Highlighting that other processes such as the direction of accretion of substructure are also major mechanisms that set the distribution of poles . Even though our sample of halos is small to draw statistical conclusions it highlights the importance of  accounting for the influence of massive satellites to interpret the present-day distribution of orbital poles.

We summarized our main findings below:

\begin{itemize}
    \item \textbf{Interactions with a massive satellite can change the distribution of orbital poles of halo tracers 
    as observed from the inner galaxy:}  
    We quantify the enhancement in apparent orbital pole clustering induced by the pericentric passage of a massive satellite using the two-point
    angular correlation function. In our calculations, the reference frame is always co-moving 
    with the central regions of the host galaxies. We find that a massive satellite on an eccentric, LMC-like orbit creates 
    changes to the apparent distribution of orbital poles of DM particles in the host galaxy halo as shown in Figures 5-7. These changes 
    are rapid and occur close to the pericentric passage of the massive satellite. This highlights the importance of accounting 
    for the dynamical state of a host galaxy and its satellites when interpreting the clustering of orbital poles of other 
    satellite galaxies.
    
    \item{\textbf{The strength of clustering induced by a massive satellite on the orbital poles of the host depends on the orbit and mass of any massive satellites:}}
    The apparent clustering of satellite orbital poles can be induced by a massive satellite. Using different metrics, like the orbital 
    poles dispersion, spherical standard deviation, and the two-point angular correlation function we found that in 
    systems with LMC-like orbits, clustering increases around the pericenter of the massive satellite.  
    Pericenter is where the COM displacement and reflex motion are strongest, as shown in Figure~\ref{fig:2d_corrfunc_alldm} 
    and \ref{fig:delta_sph}. We find that the amplitude of the clustering depends 
    on the orbit and mass of the satellite galaxy. Satellites with pericenters between 30--50 kpc, 
    high satellite--host mass ratios, and eccentricities close to 0.8 create the strongest
    enhancements to the clustering of DM particle poles (see Figure~\ref{fig:summary}). For example, 
    in the MW--LMC analog system ($m12b$) the dispersion of the orbital poles decreases by 15--20$^{\circ}$ after the first pericentric passage.

    \item{\textbf{Both the contribution from the massive satellite and the host halo response induce apparent clustering of orbital poles:}}
    We find that satellites and subhalos from the massive satellite do contribute to decreasing the dispersion of the orbital poles 
    by $\approx 5^{\circ}$ in the MW--LMC analog ($m12b$) as the satellite orbits the host. Those satellites 
    would have the same angular momentum direction as the massive satellite and hence decrease the overall 
    dispersion as shown in \cite{Samuel21}. However, we found that if those satellites are excluded, the 
    orbital poles dispersion still decreases during the pericentric passage of the massive satellite. 
    This shows that the COM displacement and reflex motion of the inner halo induced by 
    the massive satellite creates an apparent co-rotation in the outer halo (see Figure~\ref{fig:OP_disp}).

    \item{\textbf{Idealized N-body simulations agree with cosmological simulations, but not quantitatively:}}
    Idealized simulations of the MW--LMC, reproduce qualitatively the effects seen in the cosmological simulations. However, 
    the amplitude of the effects are orders of magnitude smaller than those in cosmological simulations 
    (see Figure~\ref{fig:2d_corrfunc_alldm} upper right panel). This is because 1) the phase-space distribution of satellites 
    in the cosmological halos is not captured in idealized simulations; 2) non-spherical halos as those in cosmological 
    simulations respond differently than spherical halos in equilibrium; and 3) the presence of a cosmological 
    environment can amplify the perturbations to the halo. 

    \item{\textbf{Quantitative measurements of the distribution of orbital poles depend on the halo tracer:}}    
    The correlation function measurements of star particles follow similar trends as DM particles 
    (see Figure~\ref{fig:2d_corrfunc_alldm}), but the amplitudes are higher for the star particles. 
    The mean orbital pole, on the other hand, does not necessarily always agree between satellites and 
    dark subhalos (see Figure~\ref{fig:OP_mean}), this shows that metrics like the mean are very sensitive to 
    the phase-space distribution and number of halo tracers. The dispersion in the orbital poles 
    does have similar quantitative results between satellites and subhalos when the same number of tracers are used.

    \item {\textbf{Orbital poles provide observational evidence of the dynamical state of the DM halo:}} We found that systems that experience a
    larger COM motion caused by the satellite also experience stronger and long-term changes 
    in their distribution of orbital poles. This highlights a possible connection between the amplitude of 
    the excitation of the dipole mode of the host halo with the mass and trajectory of the satellite galaxy. The massive satellites in \latte{} are on different orbits compared to the LMC and the hosts are in different 
    large-scale environments \citep{Pham22, Xu23}. These results motivate the use of the observed clustering of 
    orbital poles to constrain the orbit of the LMC and the DM halo response of the MW.

\end{itemize}

These conclusions provide insight that the pericenter passage of a massive satellite is a particular event in the 
evolution of a MW-like galaxy that impacts the observed distribution of orbital poles in the MW halo. As a consequence, placing the MW in a 
cosmological context to study co-rotation patterns must include the effect of massive satellites like the LMC close to 
the pericenter. Quantifying the probability of finding the orbital poles clustering observed in the MW in those systems 
would provide more insight into the uniqueness and formation of the VPOS.

\section{Acknowledgements}

We thank the anonymous referee for their valuable input that strengthen and help to improve the current manuscript. We are also grateful to Gurtina Besla,  Julianne Dalcanton,  and the Nearby Universe group attendees at the CCA for valuable discussions that benefited this paper. NGC thanks Jo\~ao Ant\^onio Silveira do Amarante for his guidance with pynbody. NGC thanks the SCC center and the Flatiron institute for their valuable technical support with the HPC systems.  NGC also thanks the administrative and hospitality personal at the Flatiron Institute for their continuous worked that facilitated the completion of this work. ECC acknowledges support for this work provided by NASA through the NASA Hubble Fellowship Program grant HST-HF2-51502.001-A awarded by the Space Telescope Science Institute, which is operated by the Association 
of Universities for Research in Astronomy, Inc., for NASA, under contract NAS5-26555. JS was supported by an NSF Astronomy and Astrophysics Postdoctoral Fellowship under
award AST-2102729. E.P. acknowledges support from HST GO-15902 and HST AR-16628. Support for GO-15902 and AR-16628 was provided by NASA through a grant from the Space 
Telescope Science Institute, which is operated by the Association of Universities for Research in Astronomy, Inc., under NASA contract NAS 5-26555. AW received support from: NSF via CAREER award AST-2045928 and grant AST-2107772; NASA ATP grant 80NSSC20K0513; HST grants AR-15809, GO-15902, GO-16273 from STScI.

Software: \textsc{Astropy} \citep{astropy:2018}, \textsc{Gala} \citep{gala}, \textsc{GizmoAnalysis} and \textsc{HaloAnalysis} software packages \citep{Wetzel20a, Wetzel20b}, \textsc{pygadgetreader} \citep{pygadgetreader}. \textsc{pynboddy} \cite{pynbody}, \textsc{corrfunc}\cite{corrfunc}, \textsc{scipy} \citep{scipy}.

\bibliography{references} 
\bibliographystyle{aasjournal}

\appendix

\section{Analysis for all halos in the \latte{} suite}\label{sec:appendix}

\begin{figure}[ht]
    \centering
    \includegraphics[scale=0.4]{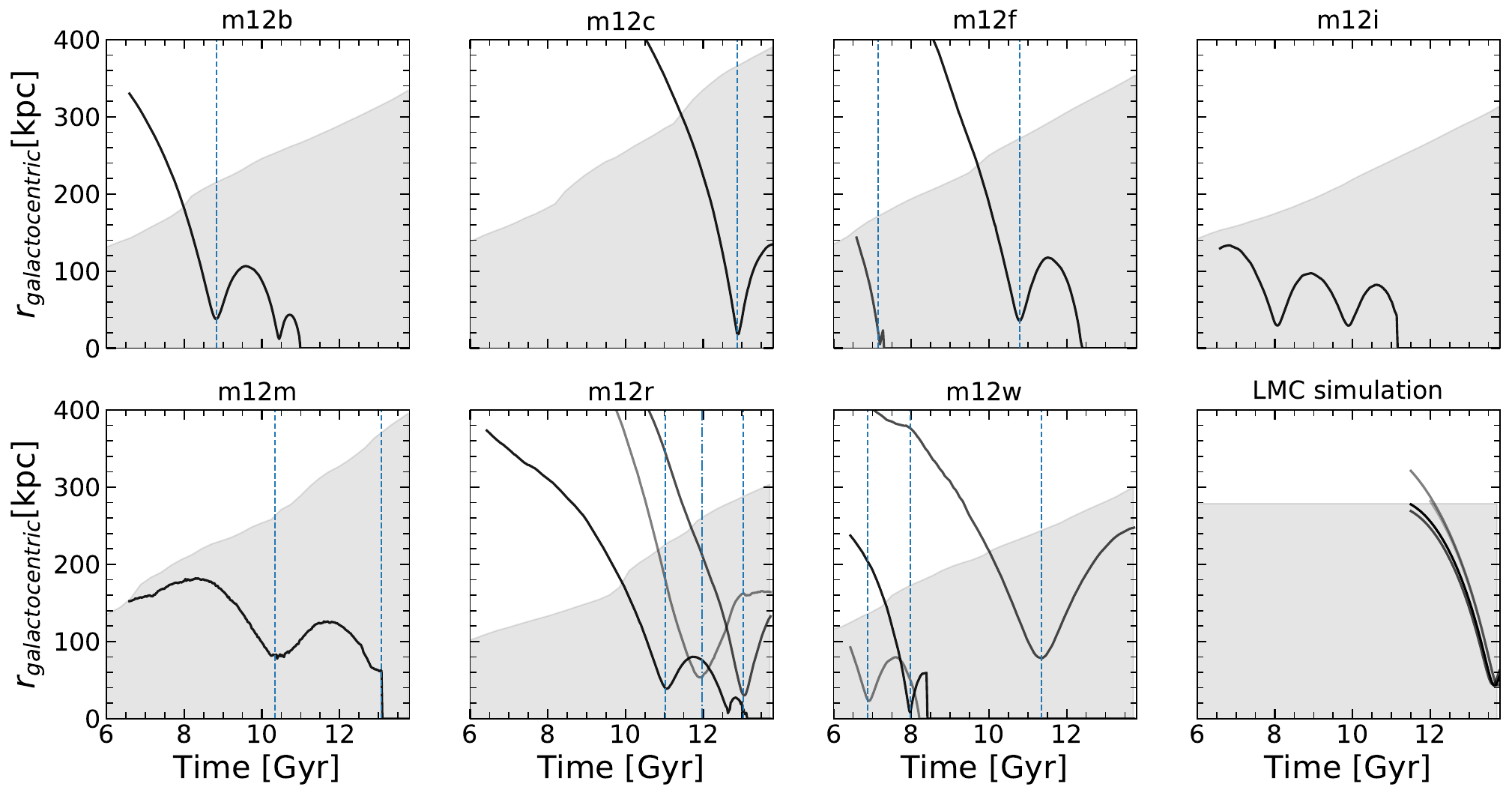}
    \includegraphics[scale=0.4]{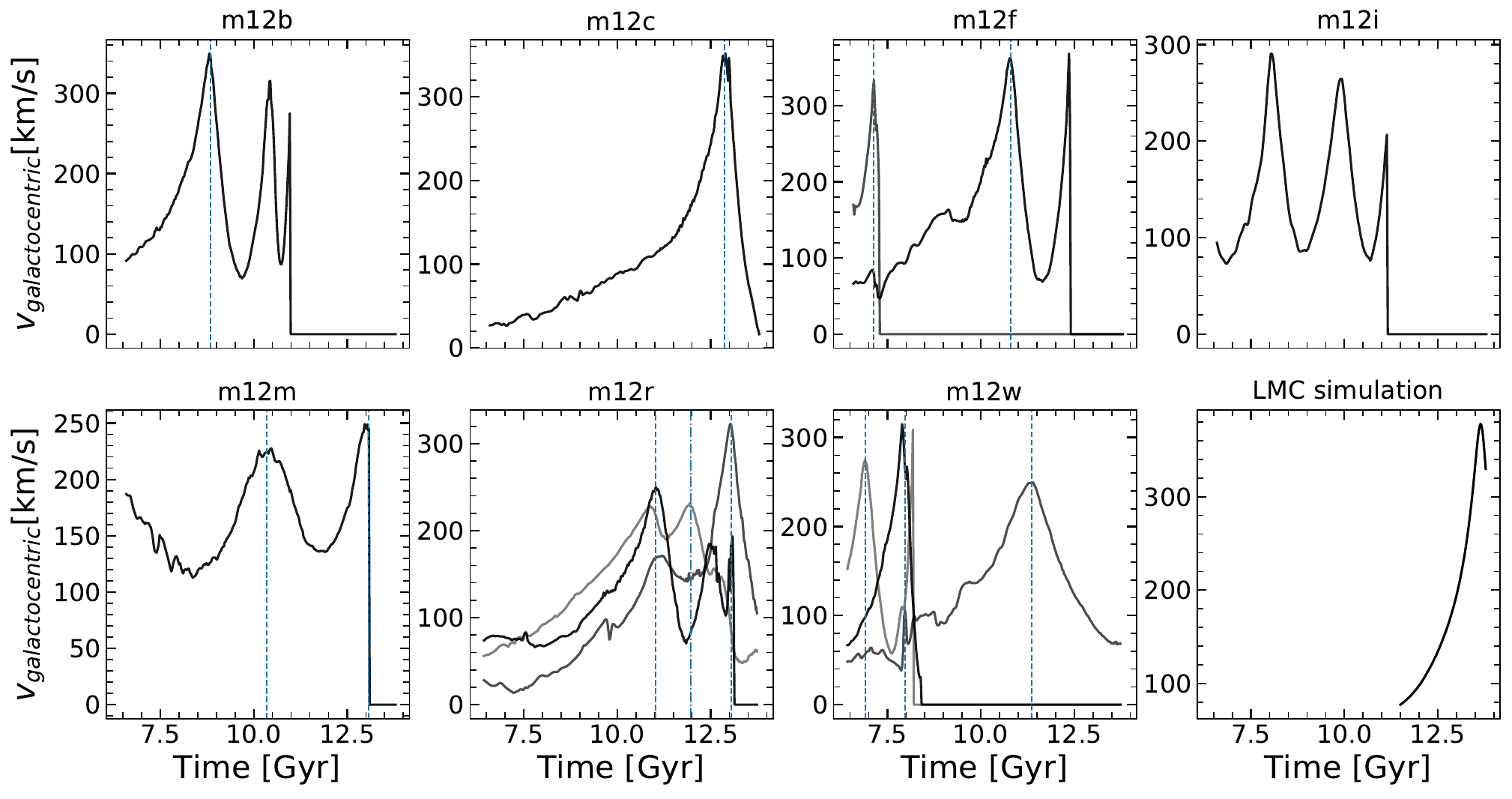}
    \caption{Galactocentric distances (upper panels) and velocities (lower panels) of the massive satellites in $m12c$, $m12f$, $m12m$, $m12r$, and $m12w$. Blue vertical lines illustrate the first pericenter passage of the massive satellites. In $m12r$, $m12w$ there are multiple satellites. The transparency of the lines are proportional to the mass of the satellites.}
    \label{fig:all_sat_orbits}
\end{figure}

Here we present the analysis for the remaining five \latte{} halos ($m12c$, $m12f$, $m12m$, $m12r$, $m12w$) as discussed in the main text. Among these five, $m12m$ does not accrete any massive satellites between $z=0-1$ and hence serves as our ``isolated halo". All the physical properties of the massive satellites and their hosts are summarized in Table~\ref{tab:fire_sats}. The pericenters, eccentricities, and satellite-halo mass ratios are shown in Figure~\ref{fig:orbits_properties} in the main text. We start by showing the orbits of the satellites in Figure~\ref{fig:all_sat_orbits}. $m12c$, $m12f$, and $m12w$ have massive satellites with pericentric distances within 20 kpc and on very radial orbits ($e>0.93$). On the other hand, $m12r$ has two massive satellites that were accreted with pericenter distances between 20-50~kpc, and are hence more likely to alter the orbital poles distribution and induce co-rotation in their hosts than satellites with larger or shorter pericenters.

\begin{figure}[ht]
    \centering
    \includegraphics[scale=0.5]{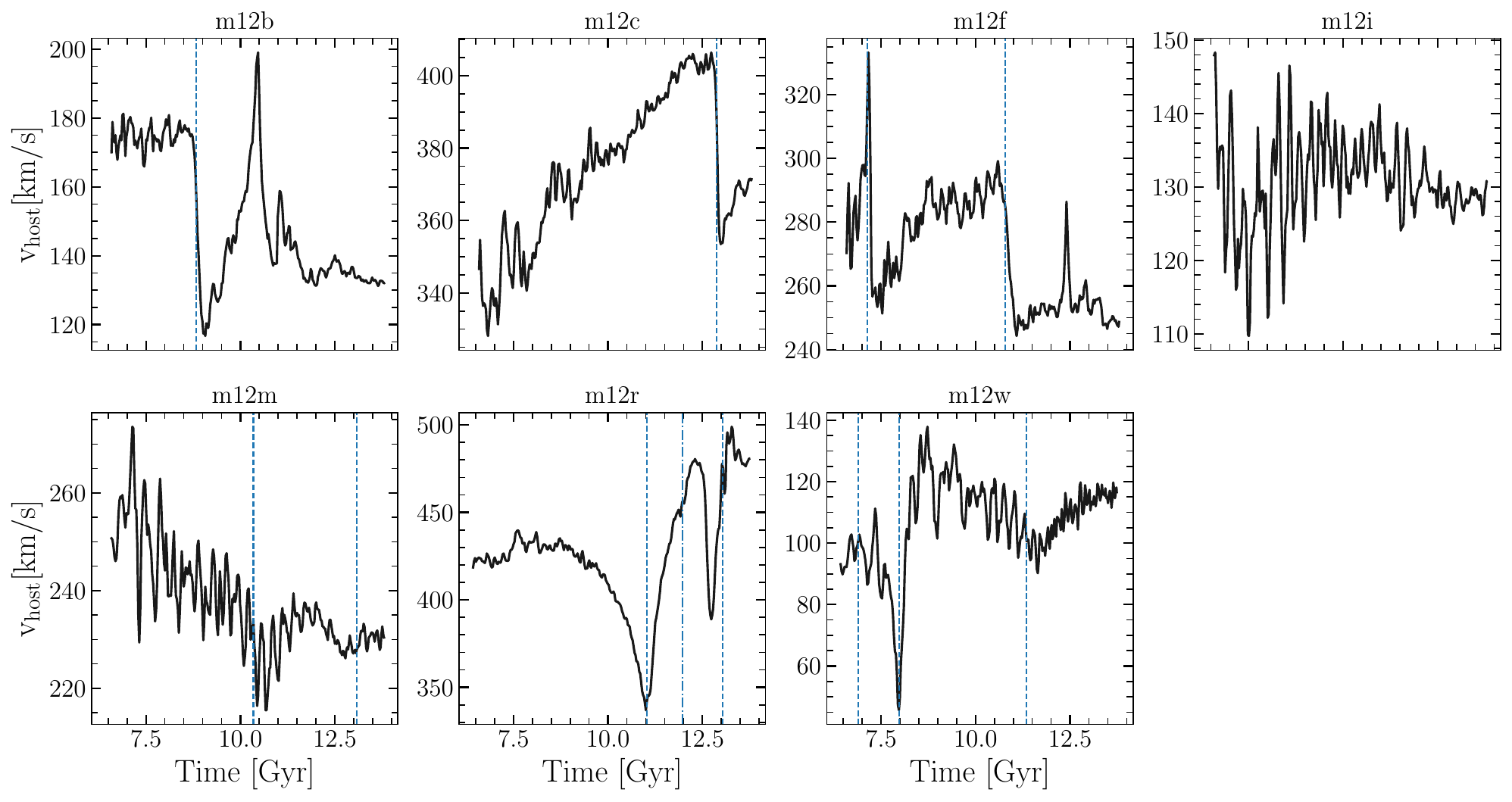}
    \caption{Velocity of the COM of the host halos as a function of cosmic time. The velocity is measured with respect to the halo's velocity at $t=6.2$ ($z=1$). Drastic velocity changes happen when massive satellites approach pericenters (indicated by the vertical blue dashed line). Changes up to 80 $\rm{km\ s^{-1}}$ can occur (for example in $m12r$) in short periods of time.}
    \label{fig:m12_vcom}
\end{figure}

In Figure~\ref{fig:m12_vcom}, we show the velocity of the host halos in the box reference frame as reported by the halo finder. For the halos with massive satellites, changes from 40~$\rm{km\ s^{-1}}$ ($m12c$, $m12f$) all the way to 80~$\rm{km\ s^{-1}}$ ($m12b$, $m12r$) are observed right after the pericentric passages of the satellites (blue dashed lines). The timescales of these changes range from 0.5--2 Gyrs. The difference in timescales and magnitudes is correlated with the pericentric distance and satellite-host mass ratios. Satellites that have pericentric distances in the range of 30--50 kpc and higher satellite-host mass induce larger changes in the host velocity and last longer.

\begin{figure*}[ht]
    \centering
    \textbf{All-sky distribution of orbital poles in the outer halo} \par\medskip
    \textbf{\ \ \ \ \  $m12c$ \ $ \ \ \ \ \ \ \ \ \ \ \ \ \ \ \ \ \ $ \ \ \ \ \ \ \ \ \ \ \ \ \  \ \ \ \ $m12f$ $ \ \ \ \ \ \ \ \ \ \ \ \ \ \ \ \ \ \ $ \ \ \ \ \ \ \ \ \ \ \ \ \ \ \ \  \ \ \ \ $m12m$} \par\medskip   
    \includegraphics[scale=0.25]{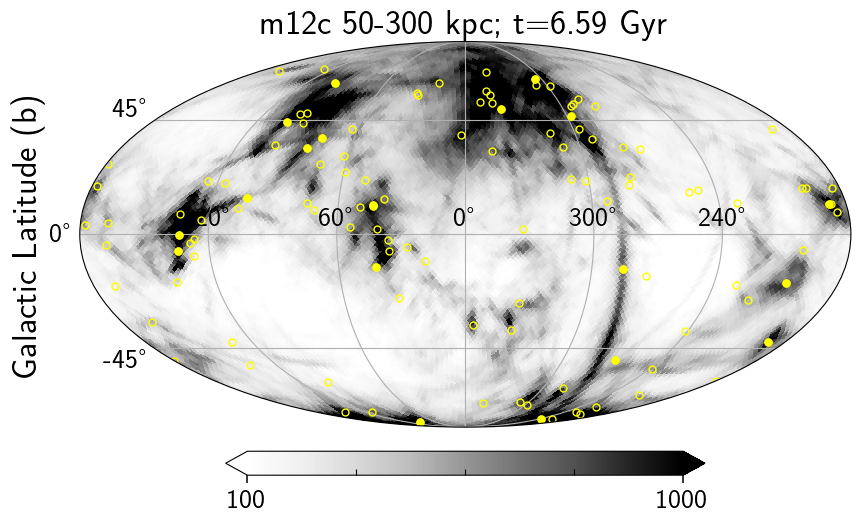}
    \includegraphics[scale=0.25]{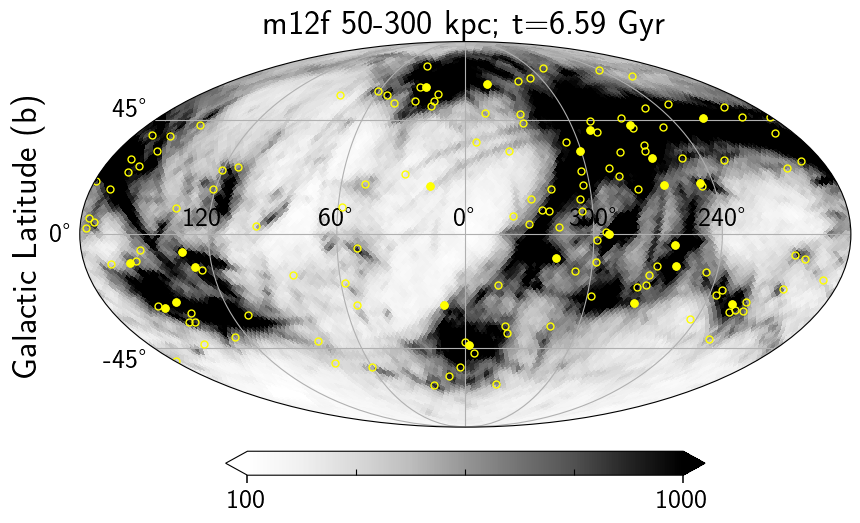}
    \includegraphics[scale=0.25]{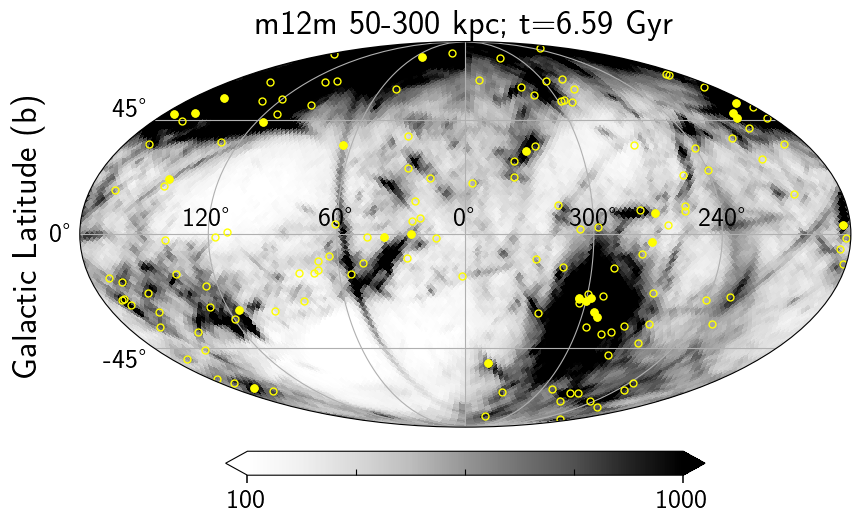}

    \includegraphics[scale=0.25]{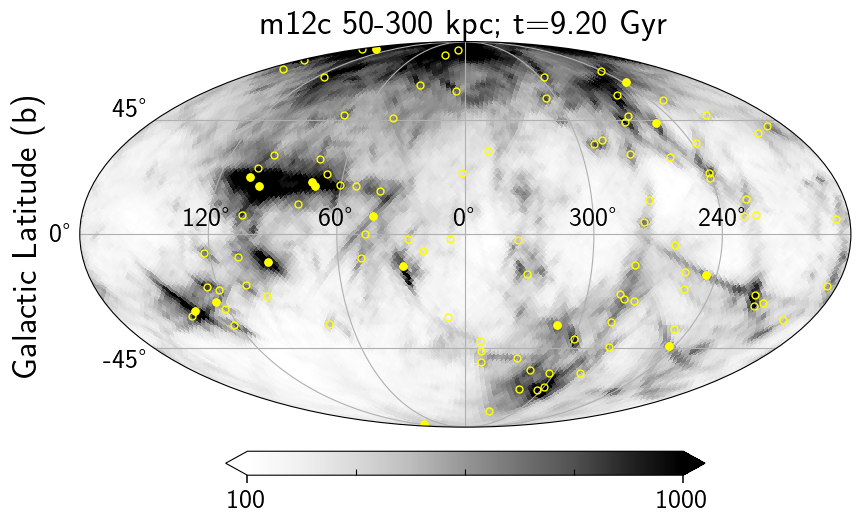}
    \includegraphics[scale=0.25]{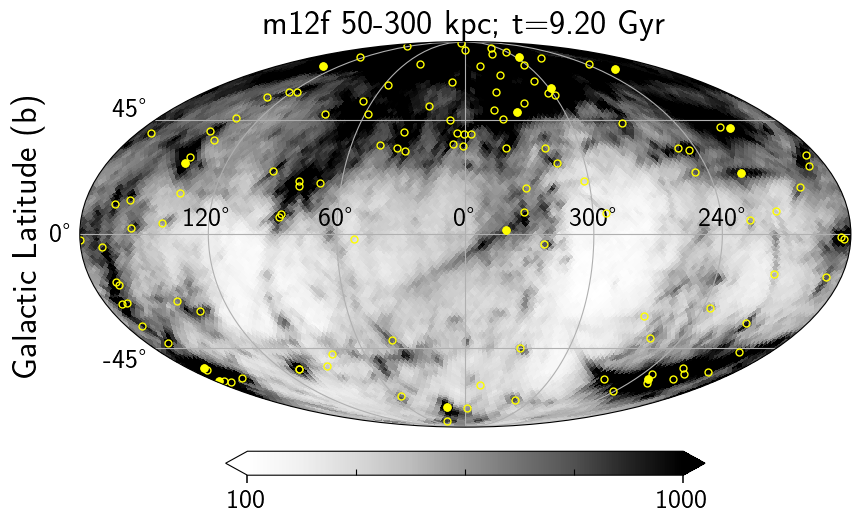}
    \includegraphics[scale=0.25]{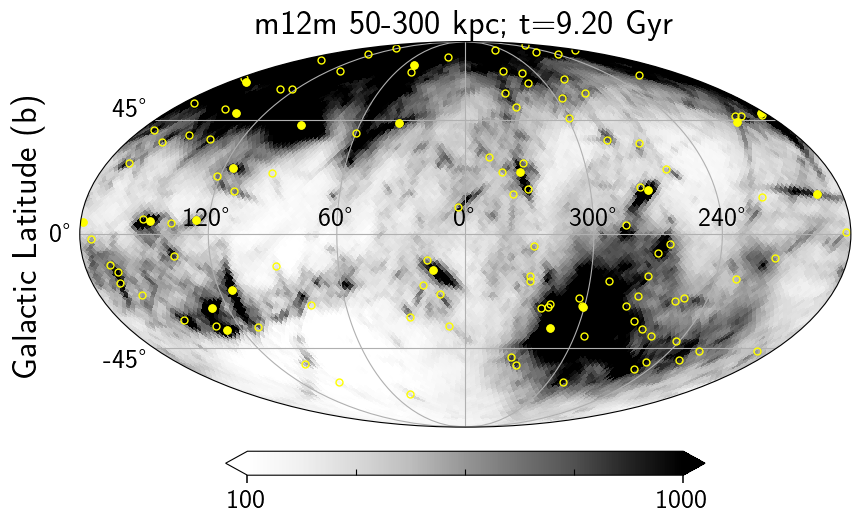}
    
    \includegraphics[scale=0.25]{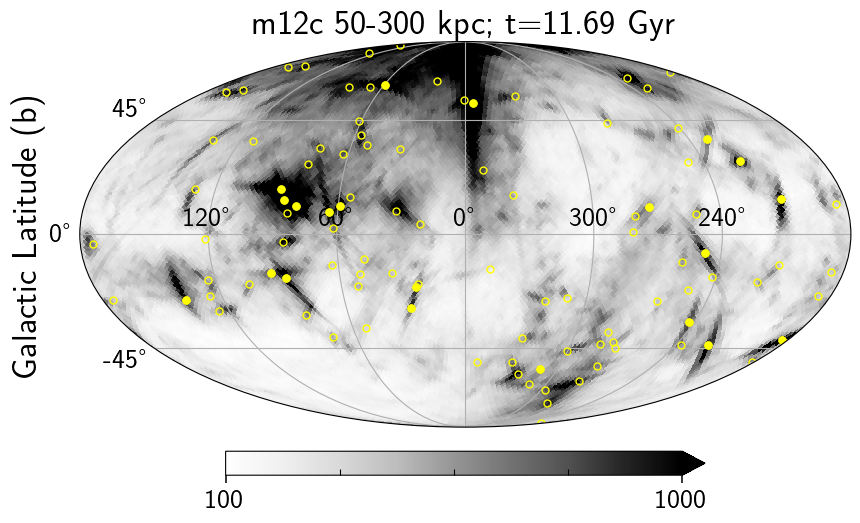}
    \includegraphics[scale=0.25]{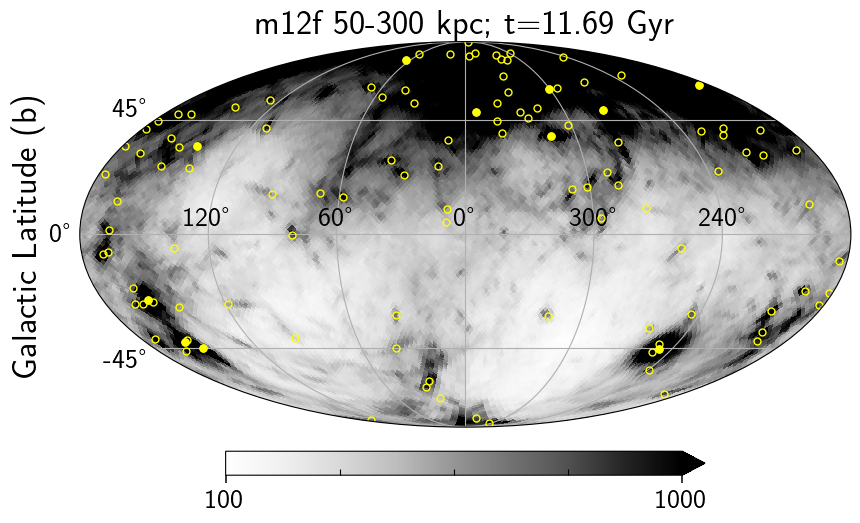}
    \includegraphics[scale=0.25]{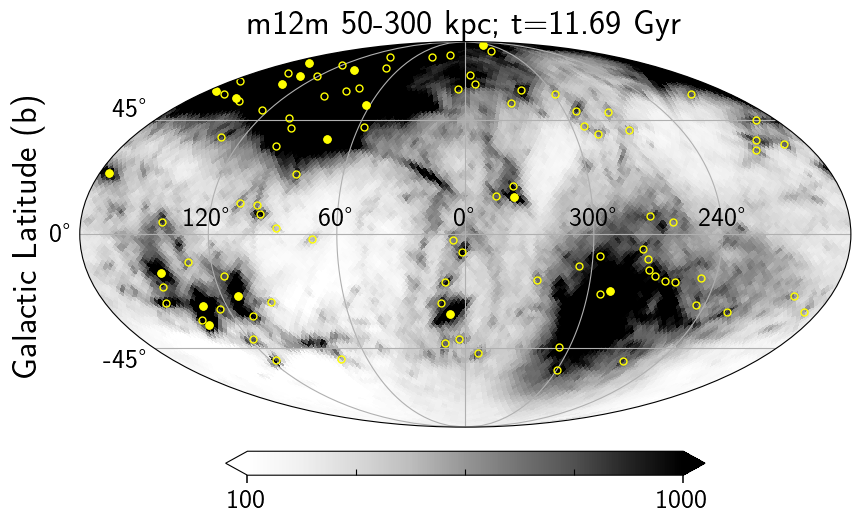}
    
    \caption{All-sky orbital poles density distribution for Dark Matter particles (Grey color map) between 
    50-300 kpc. Orbital poles of subhalos and satellite galaxies are shown with open and solid yellow
    circles respectively. Each row corresponds to a different time in the evolution of the halos and 
    each column represents a different halo $m12c$, $m12f$, $m12m$ (no massive satellite). The orbital poles 
    of the particles were computed after rotating the disk into the x--y plane and
    removing the particles from the most massive subhalos. The density of 
    orbital poles was computed using healpix with Healpy \citep{healpy}. }
    \label{fig:all_sky2}
\end{figure*}

\begin{figure*}[ht]
    \centering
    \textbf{$\ \ \ \  $ $m12r$ \ $ \ \ \ \  \ \ \ \ \ \  \ \ \ \ \ \ \ \ \ \ \ \ \ \ \ \ \ \ \ \ $ \ \ \ \ \  $m12w$} 
    \par\medskip   

    \includegraphics[scale=0.25]{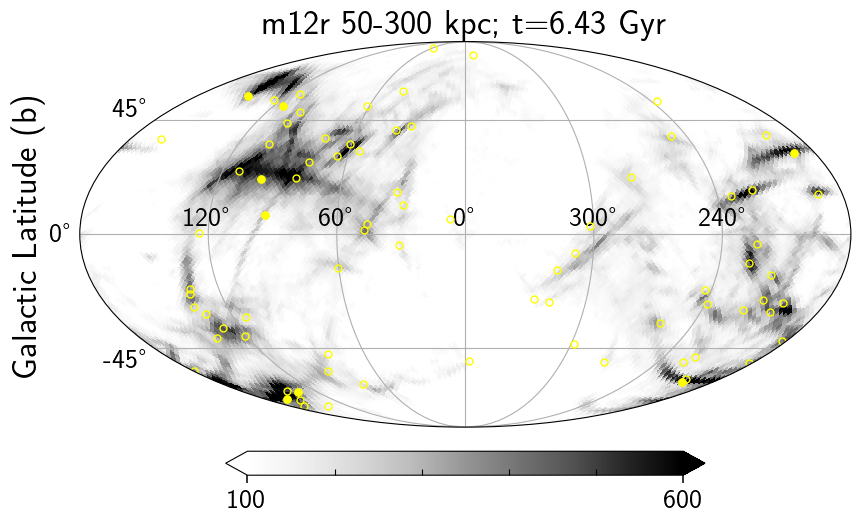}
    \includegraphics[scale=0.25]{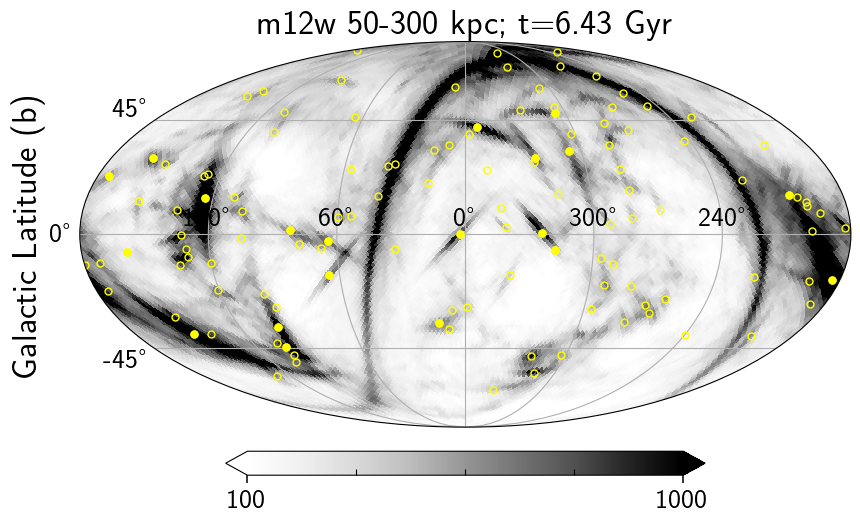}

    \includegraphics[scale=0.25]{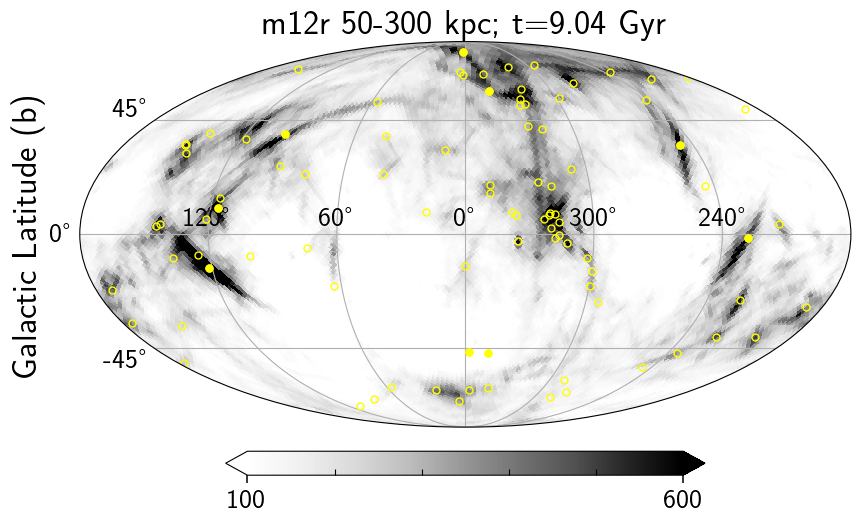}
    \includegraphics[scale=0.25]{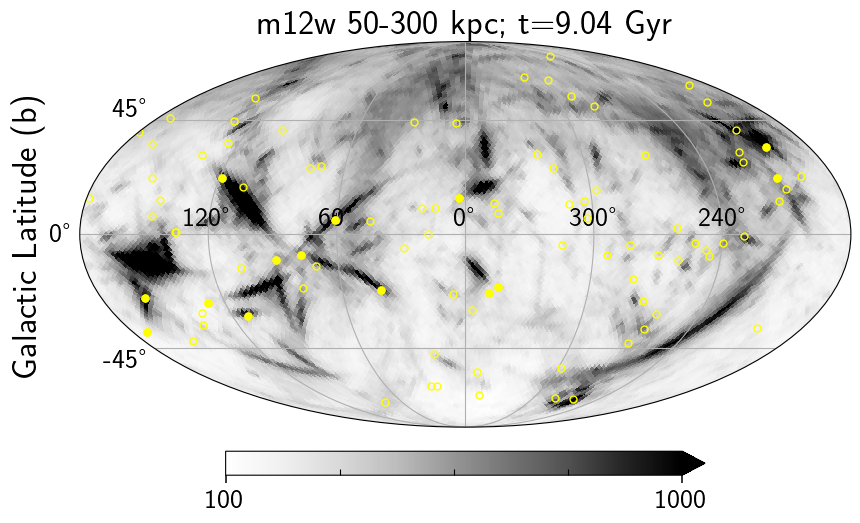}

    \includegraphics[scale=0.25]{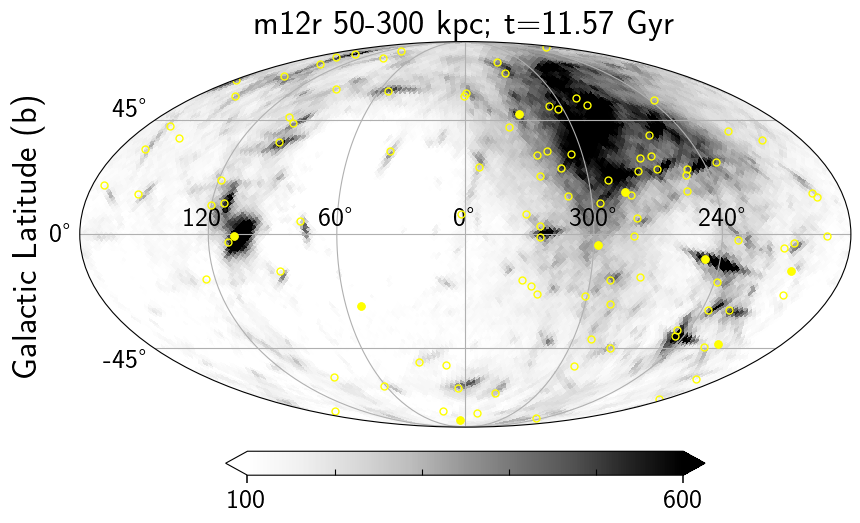}
    \includegraphics[scale=0.25]{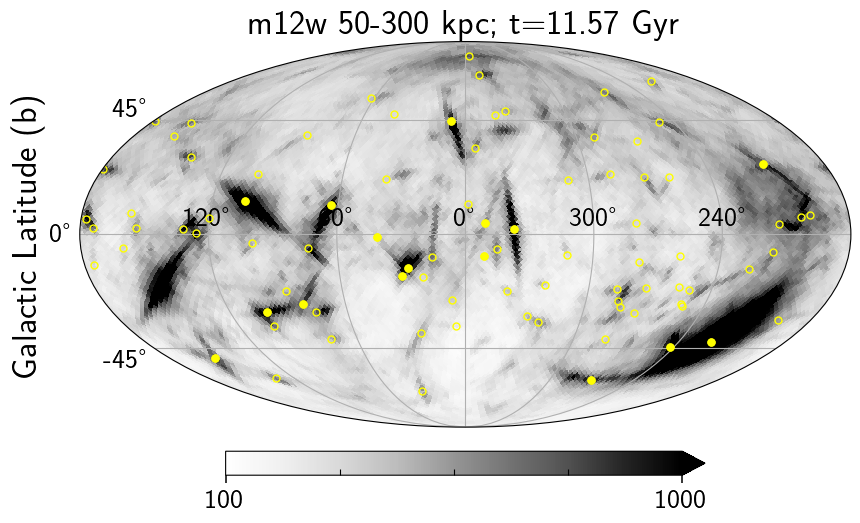}

    \caption{Same as Figure~\ref{fig:all_sky2}, but for $m12r$ (left panels), and $m12w$ (right panels). Major changes 
    in the distribution of orbital poles in $m12r$ change through the evolution of the halo. 
    Note that his halo have three major mergers after 10~Gyr, which is were most of the
     changes in the poles distribution happen. For $m12w$, the changes happen early on in the 
     distribution of poles.}
    \label{fig:all_sky3}
\end{figure*}

\begin{figure*}[ht]
    \centering
    \includegraphics[scale=0.4]{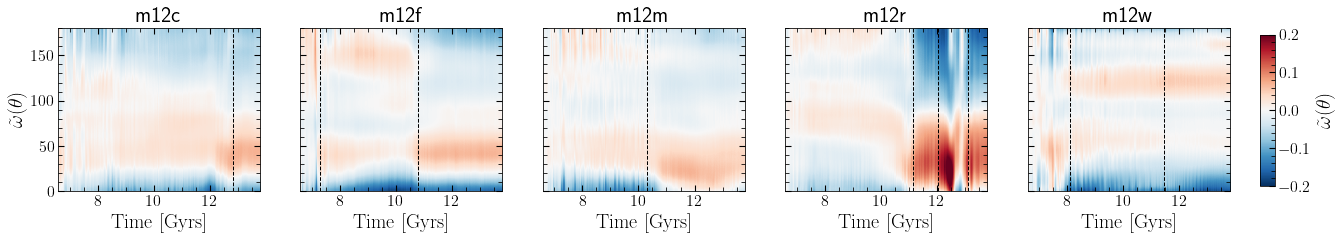}
    \caption{Temporal evolution of the correlation function for five \latte{}  halos, $m12c$ 
    (1st row), $m12f$ (2nd row), $m12m$ (3rd row), $m12r$ (4th row), and $m12r$ (5th row).  
    The correlation function is computed with $10^6$ DM particles randomly selected within 
    50-300 kpc. Red colors show an increase of probability for finding pairs at a given angular 
    distance ($\theta$). The probabilities are roughly constant time, however, when a massive 
    satellite is at pericenter the probability distribution increases at smaller angular distance.}
    \label{fig:2d_corrfunc_latte}
\end{figure*}

In Figures~\ref{fig:all_sky2} and \ref{fig:all_sky3}, we show the distribution of orbital poles of DM particles (grey color bar), subhalos (empty yellow circles), and satellite galaxies (solid yellow circles) for three different times throughout the evolution of the halos. Halos with massive subhalos exhibit rapid changes in their orbital poles distribution. In the absence of massive satellites like in $m12m$, the distribution of poles changes mildly. 
We quantify the poles distribution using the two-point correlation following the procedure described in Section~\ref{sec:corrfunc}. This is shown in Figure~\ref{fig:2d_corrfunc_latte}. In all cases, it is clear that after the pericentric passage of the massive satellites (indicated by the dashed black vertical line), the distribution of poles changes abruptly. A comparison between the degree of clustering at the angular scale where the clustering of the VPOS is observed in the MW ($\theta \approx 36^{\deg}$) is shown in Figure~\ref{fig:summary}. These results reinforce the findings presented in the main text that massive satellites do have an important impact on the distribution of orbital poles of the host.

\begin{figure*}[ht]
    \centering
    \includegraphics[scale=0.45]{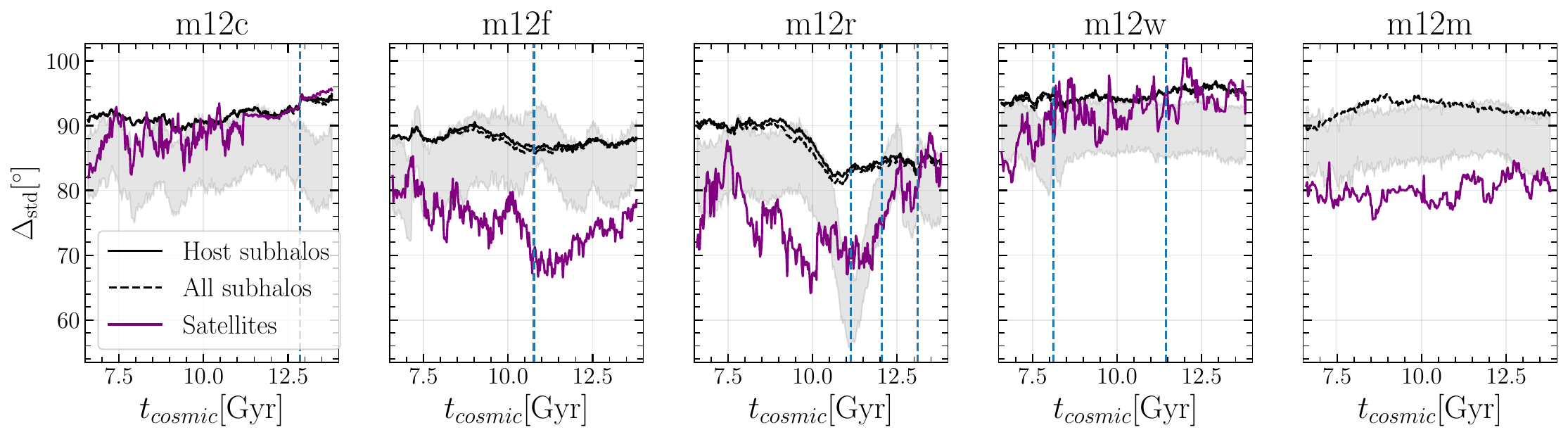}
    \caption{OThe dispersion of the orbital poles computed using subhalos with masses $\geq 10^{7}M_{\odot}$ and within 300 kpc in the 5 remaining latte halos not included in the main text. The properties of the hots halos and the massive satellites are summarized in Table~\ref{tab:fire_sats}. The grey shaded regions shows the distribution of the dispersion of the orbital poles of DM subhalos after down-sample to the same number of satellite galaxies. Blue vertical lines illustrate the first pericenter passage of the massive satellites. As also seen in the MW--LMC analog (m12b) in some cases the dispersion of the orbital poles in the host decreases close to the pericentric passage of the satellite.}
    \label{fig:op_disp_all_latte}
\end{figure*}

Similarly to Figures~\ref{fig:OP_disp}, the temporal evolution of the dispersion of the orbital poles for the remaining halos in \latte{} are shown in Figure~\ref{fig:op_disp_all_latte}. This is shown
for the subhalos (black lines) and the satellite galaxies (purple lines). Since the disprsion of the orbital poles depends on the number of subhalos or satellites, to properly compare the results between these two populations, we down-sample the subhalos to the same number of satellites. The shaded grey regions show the $1-\sigma$ measurements of the dispersion after sampling the subhalos 1000 times. We found that in $m12f$, $m12r$, and $m12m$ the dispersion of the
 orbital poles from the satellites is lower than those of the subhalos (gray shaded regions). This is because, satellites reside within the more massive subhalos which are preferentially located in the direction of accretion of the filaments. As a result, they could be more clustered and therefore have a lower dispersion of orbital poles.

Figure~\ref{fig:op_disp_all_latte} also shows that the dispersion of the orbital poles decreases when the satellite is close to its pericentric passage. In cases where the satellite's pericentric passage is between 20-60~kpc, like in $m12f$ and $m12r$, the dispersion can decrease by more than $10^{\circ}$ for the satellites.

\begin{figure*}
    \centering
    \includegraphics[scale=0.45]{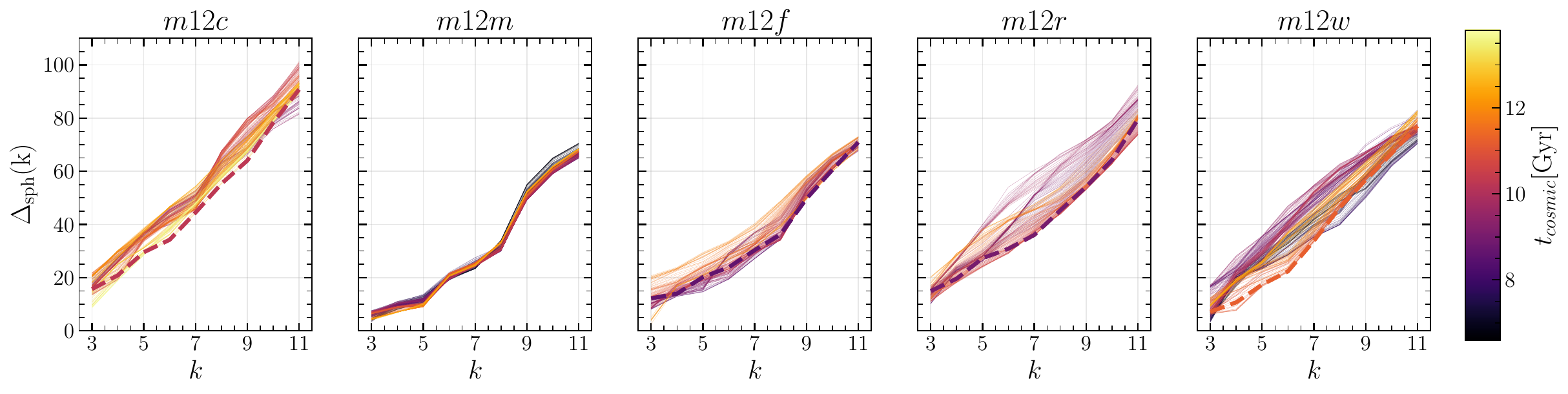}
    \caption{Temporal evolution of the spherical standard distance $\Delta_{sph}$ of the $k$ closest orbital poles of the top 11 massive satellites in the remaining halos of the \textit{latte} simulations. Smaller values of $\Delta_{sph}$ correspond to more clustered orbital poles. Dashed lines show $\Delta_{sph}$ at the pericenter passages. In halos that undergo mergers with massive satellites spread in $\Delta_{sph}$ is observed, being $m12r$ and $m12w$ the clearest cases. While in $m12m$ the evolution of $\Delta_{sph}$ is minimal.}\label{fig:dsph_all_latte}
\end{figure*}

Finally, we explore the temporal evolution of the spherical standard distance $\Delta_{sph}$ for the 11 most massive subhalos in each halo in Figure~\ref{fig:dsph_all_latte}. Here we do include the massive satellites, as this is the standard choice in the MW, where the LMC and the SMC are included when computing $\Delta_{sph}$. We found that there is a diversity of values of the $\Delta_{sph}$, $m12m$ and $m12f$ present the largest clustering for the 11 most massive satellites (lower values of $\Delta_{sph}$). In all of the halos with massive satellites we see that $\Delta_{sph}$ is minimum at the pericentric passage of the satellite (dashed line). In these halos, there is also a larger spread in $\Delta_{sph}$ highlighting that the poles of these satellites evolve more in time. Note that this is not the case for $m12m$ where the poles \textit{don't} change considerably and the spread of $\Delta_{sph}$ is the lowest among all the halos. 

\end{document}